%% file: main.tex
\documentclass[letterpaper,twocolumn,10pt]{article}
\usepackage{style}

\usepackage{hyperref}

\usepackage{xcolor}
\hypersetup{
	colorlinks=true,
	citecolor=blue,
	linkcolor=blue,
	urlcolor=blue,
}

\usepackage{cite}
\usepackage{amsmath}
\usepackage{booktabs}
\usepackage{multirow}
\usepackage{xurl}
\usepackage{amsmath}
\usepackage{amssymb}
\usepackage{amsfonts}
\usepackage{caption}
\usepackage[subrefformat=parens]{subcaption}
\usepackage{enumitem}
\usepackage{xcolor}
\usepackage{graphicx}
\usepackage{ulem}

\newcommand{\circled}[1]{\raisebox{.5pt}{\textcircled{\raisebox{-.9pt} {{\small #1}}}}}

\iftrue

\newcommand{\nsection}[1]{\section{#1}}
\newcommand{\nsubsection}[1]{\subsection{#1}}
\newcommand{\nsubsubsection}[1]{\subsubsection{#1}}

\else

\newcommand{\nsection}[1]{\vspace{-0.1in}\section{#1} \vspace{-0.03in}}
\newcommand{\nsubsection}[1]{\vspace{-0.08in}\subsection{#1}\vspace{-0.03in}}
\newcommand{\nsubsubsection}[1]{\vspace{-0.08in}\subsubsection{#1}\vspace{-0.03in}}

\fi

\usepackage{tikz}
\usepackage{amsmath}

\begin{document}

\date{}

\title{\Large \bf Trapped by Their Own Light: Deployable and
Stealth Retroreflective Patch Attacks on
Traffic Sign Recognition Systems}

\author{
{\rm Go Tsuruoka}\\
Waseda University
\and
{\rm Takami Sato}\\
University of California, Irvine
\and
{\rm Qi Alfred Chen}\\
University of California, Irvine
\and
{\rm Kazuki Nomoto}\\
Waseda University
\and
{\rm Yuna Tanaka}\\
Waseda University
\and
{\rm Ryunosuke Kobayashi}\\
Waseda University
\and
{\rm Tatsuya Mori}\\
Waseda University, RIKEN AIP, NICT
}

\maketitle

\input{src/00-Abstract}
\input{src/01-introduction}
\input{src/02-Background-RelatedWork}

\input{src/03-Methodology}
\input{src/032-modeling_attack_optimization}
\input{src/04-Evaluation1}

\input{src/05-Evaluation2.1}

\input{src/05-Evaluation2.2-Stealthiness}

\input{src/06-DrivingEvaluation.tex}

\input{src/07-Defence}
\input{src/08-Discussion_Limitaiton}

\input{src/09-Conclusion}

\bibliographystyle{IEEEtran}
\bibliography{ref}

\input{src/10-Appendix}
\end{document}

%% file: src/00-Abstract.tex
\begin{abstract}
\nocite{DBLP:journals/corr/abs-1807-07769}
\nocite{zhao2018seeing}
\nocite{sato2024invisible}
\nocite{SLAP}
Traffic sign recognition plays a critical role in ensuring safe and efficient transportation of autonomous vehicles but remain vulnerable to adversarial attacks using stickers or laser projections. 
While existing attack vectors demonstrate security concerns, they suffer from visual detectability or implementation constraints, suggesting unexplored vulnerability surfaces in TSR systems. We introduce the Adversarial Retroreflective Patch (ARP), a novel attack vector that combines the high deployability of patch attacks with the stealthiness of laser projections by utilizing retroreflective materials activated only under victim headlight illumination.
We develop a retroreflection simulation method and employ black-box optimization to maximize attack effectiveness. ARP achieves $\geq$93.4\% success rate in dynamic scenarios at 35 meters and $\geq$60\% success rate against commercial TSR systems in real-world conditions. Our user study demonstrates that ARP attacks maintain near-identical stealthiness to benign signs while achieving $\geq$1.9× higher stealthiness scores than previous patch attacks. We propose the DPR Shield defense, employing strategically placed polarized filters, which achieves $\geq$75\% defense success rates for stop signs and speed limit signs against micro-prism patches.

\end{abstract}

%% file: src/01-Introduction.tex
\nsection{Introduction} \label{sec:intro}

Traffic signs are fundamental to road safety and critical for autonomous driving (AD) systems to prevent accidents. Vision-based Traffic Sign Recognition (TSR) systems~\cite{Vision-Based-Traffic-sign-recognition} are widely deployed in modern vehicles, including Tesla Autopilot~\cite{Tesla-autopilot} and Toyota Road Sign Assist~\cite{Toyota-road-sign-assist}. Tesla's Full Self-Driving (FSD)~\cite{tesla_fsd_manual} integrates TSR for autonomous stopping at stop signs and speed limit compliance.
    However, recent lines of research~\cite{adv-patch, DBLP:journals/corr/abs-1807-07769, SLAP, wang2023rfla,sato2024invisible} have actively reported the potential vulnerability of vision-based TSR systems against adversarial attacks. Patch attacks~\cite{adv-patch,DBLP:journals/corr/abs-1807-07769} and light or laser projection attacks~\cite{phantom,SLAP,wang2023rfla,sato2024invisible} are the two major attack vectors, but we identify that these attacks still suffer from major limitations in the attack stealthiness and deployability in the real world:\textit{Patch attacks}~\cite{adv-patch,DBLP:journals/corr/abs-1807-07769} utilize small physical stickers on traffic signs to fool TSR systems. Although designed to be small and stealthy, humans can clearly see them and pedestrians or road guards may immediately remove them. \textit{Light or laser projection attacks}~\cite{phantom,SLAP,wang2023rfla,sato2024invisible} achieve high stealthiness as attackers can selectively turn on the projection only during intended attacks, but require large equipment and attacker presence within 25~m of the target sign~\cite{sato2024invisible}, which falls within the standard visibility range (100~m) established by FHWA design guidelines~\cite{MUTCD2023_2A}, reducing deployability.

\begin{figure}[t!]
    \centering
     \includegraphics[width=\linewidth]{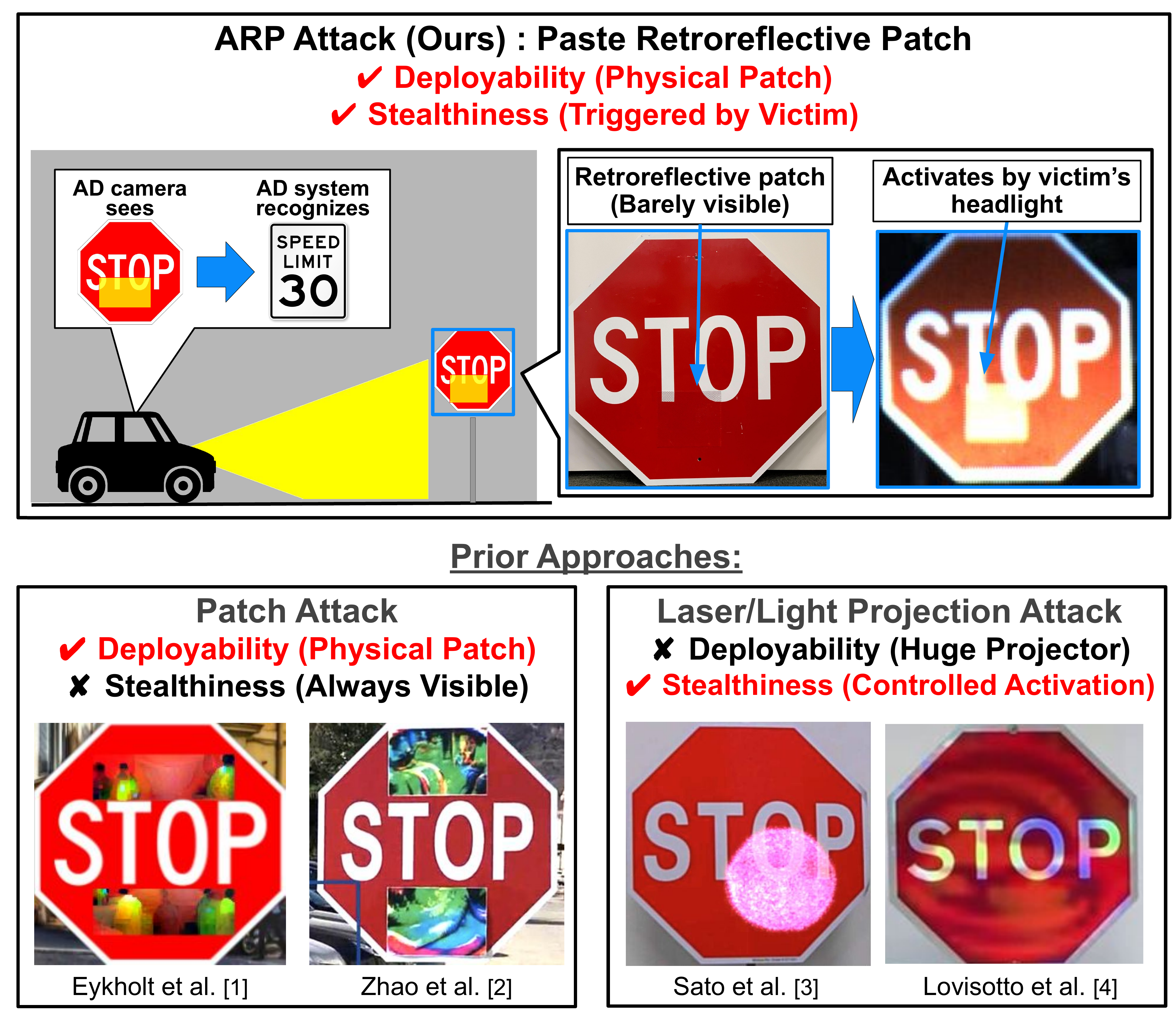}

    \caption{Overview of ARP Attack. The attack patch is triggered only when the victim's headlights shine it. This design allows to achieve as high deployability as patch attacks and as high stealthiness as light/laser projection attacks.
    }
    \label{fig:poc}
    \vspace{-5mm}
\end{figure}

To address the limitations, we introduce \textit{Adversarial Retroreflective Patch (ARP)} attacks that combine the advantages of both approaches. ARP exploits retroreflective materials—which reflect light back to their source—to create patches triggered only by victim headlights. Our patches match target sign colors (e.g., white patches for white areas), appearing as natural extensions of original signs. This achieves dual benefits: daytime stealth through color-matching and material similarity, and nighttime activation via headlight illumination. Unlike projection attacks, ARP maintains stealthiness to non-target observers due to retroreflective materials' directional selectivity.

Realizing effective ARP attacks presents two key technical challenges that prior methodologies cannot address: 
\textbf{(1) Lack of accurate modeling of retroreflective reflection} prevents us from adequately assessing the threat of ARP attacks. 

Prior approaches~\cite{DBLP:journals/corr/abs-1807-07769, zhao2018seeing, jia2022foolingeyesautonomousvehicles} only model color changes in adversarial patches. However, retroreflective reflection involves complex optical phenomena that cannot be accurately reproduced without proper modeling of underlying optical mechanisms. The reflection color is determined by intricate material properties including reflection coefficient, base color, and multi-layer structure, as well as incident light conditions. Even state-of-the-art 3D rendering simulators like Blender do not natively support retroreflective reflection simulation, highlighting the complexity of accurately modeling these physical phenomena.
\textbf{(2) Lack of methodology to effectively generate effective and stealthy ARP attacks whose color transit according to incident lights}. 
Prior attack generation methodologies ~\cite{DBLP:journals/corr/abs-1807-07769, zhao2018seeing, jia2022foolingeyesautonomousvehicles} presume that the colors on the attack patch will not be changed as their attack vectors are stickers or paints. As described in~\S\ref{sec:retro_material}, the background color of the retroreflective patch can be flexibly chosen by the attacker. Ideally, the attacker can use almost the same color as the background color where the patch is placed if such colors are available in the market, as shown in Fig.~\ref{fig:poc}. 

As a trade-off, the resulting retroreflective color is non-trivially determined by complex material properties and optical phenomena, limiting attackers' direct control.
These key advantages and limitations of ARP attacks cannot be handled by prior approaches.

This study is motivated to address these two limitations of previous studies and design a methodology to investigate effective and stealthy ARP attacks to properly evaluate their security impact on AD. In~\S\ref{sec:threat}, we first formulate our ARP threat model and investigate the physical and optical characteristics of the retroreflective materials through the measurements of the four different retroreflective materials covering the major grades defined in industrial standards. In \S\ref{sec:methodology}, we develop a methodology to accurately model retroreflective reflection based on our measurement results and design a novel attack generation pipeline to find effective and stealthy ARP attacks. In~\S\ref{sec:evaluation}, we evaluate the ARP attack effectiveness in the digital space and the physical world. We also evaluate the stealthiness via a user study showing that ARP attacks are perceived as natural as benign signs (avg. score 2.04 vs 1.81 on a 5-point scale), while prior attacks are consistently rated as unnatural (avg. score $\geq$3.69).

In~\S\ref{sec:driving}, we evaluate the ARP attack on Toyota Yaris 2024 and Nissan E-Note 2024 and confirm that the ARP attack can achieve $\geq$60\% attack success rates (ASR) even against commercial TSR systems.
In~\S\ref{sec:defense}, we design an effective defense, named DPR Shield, that leverages the characteristics of retroreflective reflection. We utilize a pair of polarized filters, one placed on the camera and another one placed on the light source. We find that DPR Shield can achieve  $\geq$75\% defense success rates for stop signs and speed limit signs. Finally, we discuss the implications and limitations of this study in~\S\ref{sec:discuss}.

    In summary, our study makes the following contributions:
\begin{itemize}[leftmargin=0.12in]
\vspace{-0.09in}
\setlength{\parskip}{0.05mm} 
\item We introduce Adversarial Retroreflective Patch (ARP) attack, a novel attack vector that exploits retroreflective materials to achieve high daytime stealthiness while maintaining nighttime effectiveness, combining the deployability of patch attacks with the stealth benefits of projection attacks.

\item We address key technical challenges by developing accurate retroreflective reflection modeling based on physical characterization of four industrial-grade materials, and designing a black-box optimization pipeline that balances stealth and effectiveness.

\item We demonstrate comprehensive attack effectiveness across multiple scenarios: $\geq$93.4\% attack success rate at $\geq$35m distance in simulation environments, and $\geq$60\% against Toyota Yaris 2024 and Nissan E-Note 2024.

\item We validate attack stealthiness via user studies, showing ARP attacks achieve near-natural appearance (avg. score 2.04 vs 1.81 for benign signs) while prior patch attacks are consistently detected as unnatural (avg. score $\geq$3.69).

\item We design DPR Shield, a physics-based defense mechanism using polarized filters that leverages retroreflective material properties to achieve $\geq$75\% defense success rates for stop signs and speed limit signs without computational overhead.

Demo videos are available at~\url{https://sites.google.com/view/tsr-retroreflectivec-attack/}.

\end{itemize}

%% file: src/02-Background-RelatedWork.tex
\vspace{-0.15in}
\nsection{Background and Related Work}
\label{sec:background}

\nsubsection{Vision-Based Traffic Sign Recognition}

    Vision-based Traffic Sign Recognition (TSR) systems have two major types of architectures as also discussed in~\cite{sato2024invisible}: \textit{single-stage} and \textit{two-stage}:

    \textit{Single-stage TSR} utilizes a single DNN-based object detector, such as YOLO series~\cite{YOLO}, not only to localize the position of traffic signs in the input images but also to classify the types of the detected traffic signs with a multi-class classification head. The single-stage TSR has clear advantages in design simplicity and computational cost, but the single-stage TSR is known to have a limitation in the scalability to handle a large number of traffic sign classes and does not yield acceptable performance with the 314 classes~\cite{mapillary_dataset}. Thus, single-stage TSR is typically adopted by up to Level-2 AD which does not need to handle a wide variety of traffic signs.

    \textit{Two-stage TSR} adopts two DNN models, object detectors and classifiers, to handle the two different tasks: localization and classification, respectively. The first-stage object detector crops the input image where the detected traffic signs exist. The second-stage classifier classifies the cropped region in the first stage. By splitting the task into 2 subtasks (cropping and classifying), the two-stage TSR can so far handle more traffic sign classes than the single-stage TSR~\cite{sato2024invisible}.
    In this work, we investigate the attack impact on both architectures.

\nsubsection{Adversarial Attacks on TSR systems}

    After adversarial attacks~\cite{Szegedy2014, goodfellow2014explaining} have demonstrated their significant performance to compromise DNN models, recent research has shown their further attack capabilities and realizability in the physical world~\cite{kurakin2016adversarial_b, sharif2016accessorize, athalye2018synthesizing, brown2017advpatch, chen2018shapeshifter, adv-patch,  zhao2018seeing, pei2017deepxplore, tian2018deeptest, chernikova2019self, zhou2018deepbillboard}.
    Patch attacks~\cite{adv-patch,DBLP:journals/corr/abs-1807-07769} and light or laser projection attacks~\cite{phantom,SLAP,wang2023rfla,sato2024invisible} have emerged as the two major attack vectors against TSR systems.
    Patch attacks~\cite{adv-patch,DBLP:journals/corr/abs-1807-07769} have shown that small benign-looking stickers can fool the detection and classification. 
    Light or laser projection attacks~\cite{phantom,SLAP,wang2023rfla,sato2024invisible} have demonstrated that light or laser traces can be sufficient to fool the TSR systems. However, the patch attacks have limitations in their stealthiness due to the visibility of the stickers that are permanently placed. The light or laser projection attacks have limitations in the attack deployability due to their large attack device to project the target sign. Our ARP attack is designed to achieve the advantages of both methods. The retroreflective patches are easy to deploy and keep stealthiness other than the attack time when the victim vehicle shines the stickers with its headlights.
    We note that Chen et al.~\cite{chen2024reflective} treats reflective materials as uniform white patches for pedestrian detection attacks, lacking consideration of physical phenomena and material properties while relying on simplified digital simulations without physical validation, thus representing a straightforward extension of conventional patch attacks that was not our comparative focus.

\nsubsection{Retroreflective Materials} \label{sec:retro_material}

    Retroreflection is a type of reflection that reflects light back towards the light source. This phenomenon is enabled by specially designed surfaces or substances with glass beads or prismatic elements to minimize scattering. One of its critical applications is traffic signs, where retroreflective materials enable high visibility even at night by efficiently redirecting vehicle headlight beams back to drivers.

\begin{figure}[!t]
    \centering
\includegraphics[width=\linewidth]{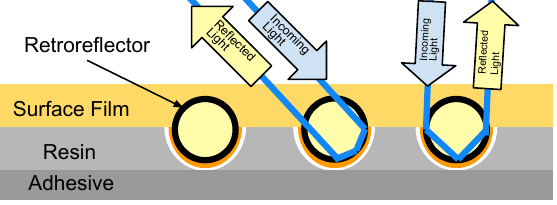}

    \caption{
    Structure of a retroreflective sheet. The surface film determines the color, while the inner retroreflector layer (glass beads or micro-prisms) in charge of the retroreflectivity, which reflect incoming lights back to their source direction as drawn with the light path diagrams.
    }
    \label{fig:retroreflective-sheeting-structure}
    \vspace{-0.2in}
\end{figure}

    The structure of retroreflective materials consists of multiple functional layers, as illustrated in Fig.~\ref{fig:retroreflective-sheeting-structure}. The bottom adhesive layer secures the material to the sign surface, while a resin layer above it holds the retroreflector - the key component that enables retroreflective functionality. Two major types of retroreflectors exist: glass beads and micro prisms~\cite{fhwa-traffic-sign-retroreflective-sheeting}.
    
    Glass bead-based retroreflectors operate through a combination of refraction and reflection, where light enters the bead, refracts, and reflects back from the material behind it. While glass bead-based retroreflectors are economical, they offer inferior performance compared to micro prism-based retroreflectors. Micro-prism retroreflectors have become the industry standard due to their superior reflection characteristics. Our market analysis of the top 50 ``retroreflective sheeting'' products on Amazon~\cite{amazon_website} found 28 products using micro-prism technology, while none used glass beads. The remaining 22 products used non-standard plastic-based materials.
    
    The performance of retroreflective materials is quantified by the coefficient of retroreflection, which represents the ratio of luminance from the reflective surface to incident illuminance. This coefficient is regulated by two major standards: ASTM (American Society for Testing and Materials)~\cite{ASTM-D4956-19} and AASHTO (American Association of State Highway and Transportation Officials)~\cite{Hawkins2022}. While ASTM E810 provides detailed laboratory testing protocols, AASHTO M 268 focuses on practical performance for road applications. As listed in Table~\ref{tbl:retro_standards}, under these standards, glass-based retroreflectors only achieve Grade III in ASTM D4956 and Grade A in AASHTO - performance levels not recommended for traffic sign use~\cite{moeur2019manual}. In contrast, high-performance prismatic materials consistently achieve retroreflection coefficients that meet or exceed the requirements for traffic sign applications.

\begin{table*}[t!]
\centering
\footnotesize
\renewcommand{\arraystretch}{1.1}
\setlength{\tabcolsep}{1pt}
\caption{Grade Taxonomy and typical applications of retroreflective materials in ASTM D4956~\cite{ASTM-D4956-19} and AASHTO~\cite{Hawkins2022} standards.}

\begin{tabular}{|c|cc|l|}
\hline
\multirow{2}{*}{Materials}    & \multicolumn{2}{c|}{Sheeting Grades}                  & \multicolumn{1}{c|}{\multirow{2}{*}{Typical Uses~\cite{moeur2019manual}}}                     \\ \cline{2-3}
                              & \multicolumn{1}{c|}{ASTM} & AASHTO             & \multicolumn{1}{c|}{}                                                  \\ \hline \hline
\multirow{3}{*}{Glass beads}  & \multicolumn{1}{c|}{I}          & N/A                & \multirow{2}{*}{Vehicle graphics, Advertising signs, and Conspicuous markings (This type is no longer in general use for traffic signs)}                   \\ \cline{2-3}
                              & \multicolumn{1}{c|}{II}         & N/A                &                                                                        \\ \cline{2-4} 
                              & \multicolumn{1}{c|}{III}        & A                  & Sperseded by newer alternatives such as Type IV (Not recommended for traffic signs)                       \\ \hline
\multirow{4}{*}{\begin{tabular}[c]{@{}c@{}}Micro Prisms \\ (Traffic Sign Grade)\end{tabular}} & \multicolumn{1}{c|}{IV}         & \multirow{3}{*}{B} & Post-mounted signs, barricades, and rigid temporary traffic control devices     \\ \cline{2-2} \cline{4-4} 
                              & \multicolumn{1}{c|}{VIII}       &                    & Post-mounted and overhead signs, internally-illuminated signs, barricades and semi-rigid temporary devices      \\ \cline{2-2} \cline{4-4} 
                              & \multicolumn{1}{c|}{IX}         &                    & Post mounted signs, overhead signs, Internally-illuminated signs, and signs with flourescent backgrounds \\ \cline{2-4} 
                              & \multicolumn{1}{c|}{XI}         & D                  & Comprehensive use for diverse sign types including fluorescent backgrounds \& rigid temporary traffic control devices          \\ \hline
\end{tabular}
\label{tbl:retro_standards}
\vspace{-0.1in}
\end{table*}

%% file: src/03-Methodology.tex
\nsection{Threat Model and Attack Capabilities}
\label{sec:threat}

\nsubsection{Threat Model}\label{subsec:threatmodel}

\begin{figure}[t]
        \centering
        \includegraphics[width=1\linewidth]{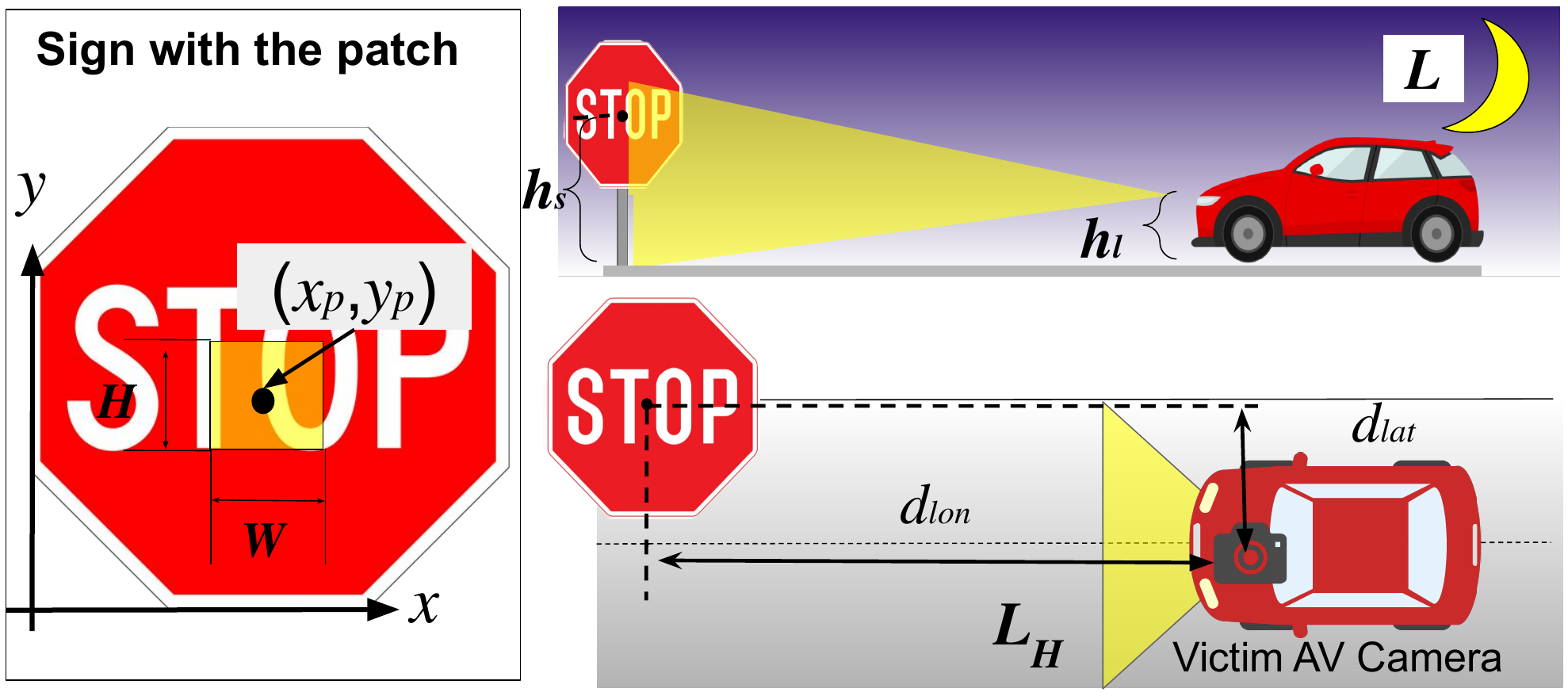}
        \caption{Overview of the ARP attack threat model. The victim's headlights are considered as the attack trigger, and parameters are adopted from previous patch-based attacks.}
        \label{fig:threat_model}
    \end{figure}

    Fig.~\ref{fig:threat_model} illustrates an overview of the ARP attack threat model, which generally follows the same threat model as conventional patch attacks. The major difference from the prior patch attacks is the prerequisite of the victim vehicle's headlights to trigger the attack. The definitions of the attack and scenario parameters are listed in Table~\ref{Table:thread_model_parameters}.
    The attacker can place malicious retroreflective patches indexed by $i$ at $(x_p^i, y_p^i)$ with the size of $(W^i, H^i)$ on the target traffic signs. To maintain attack stealthiness while ensuring effectiveness, we constrain the total area of each patch ($W^i \times H^i$) to be at most $MPR$ of the target sign's bounding box area ($W_{bbox} \times H_{bbox}$). For example, with a speed limit sign measuring 24 inches × 30 inches (720 square inches), a $MPR$ of 0.0625 corresponds to 45 square inches. The base colors of the retroreflective patches ($C^i$) must be the same color as the surface they are applied to be stealthy as shown in Fig.~\ref{fig:four_patch_day_night} in Appendix~\ref{sec:appendix-patch-day-night}.
    We thus assume the attack time is mainly during dark hours from evening to early morning.
    For this model, we define the height of the sign ($h_s$), the height of the headlight ($h_l$), the intensity of ambient light ($L$), and the intensity of the headlight ($L_{H}$). We assume that the attacker can estimate or know these scenario parameters by observing the victim and environment.
    
    The attacker's objective is to cause serious traffic rule violations that could be dangerous or even fatal by fooling the vision-based TSR systems in AD vehicles. For example, misdetection of a stop sign could lead to a crash with other vehicles at an intersection; misclassification of speed limit signs could cause unexpected acceleration or deceleration.

    \nsubsection{Retroreflective Material Selection} \label{sec:patch_selection}
    
    To comprehensively and systematically investigate the capability of the ARP attack, we selected four representative retroreflective materials for this study as listed in Table~\ref{table:selected_patch}. The selected patches are chosen to widely cover different grades and structures of retroreflective materials. 
    Our selection covers major currently available ASTM D4956 grades (Type III, IV, VIII, and XI) and all AASHTO micro-prismatic grades (B through D) used for traffic signs as shown in Table~\ref{tbl:retro_standards} and selected one glass-bead retroreflective material.

    As listed in Table~\ref{tbl:retro_standards}, our selection includes one glass-bead-based material representing entry-level technology and covers major currently available ASTM D4956 grades above III and all AASHTO micro-prismatic grades (B through D) used for traffic signs.    
    We also selected the brands that can cover the colors used in the US-style traffic signs, especially for the stop and speed limit signs, such as red, white, and black.
    The NittoL~\cite{NittoWideReflect} is a glass bead-based retroreflective material, which is cheap (80 USD per $\text{m}^2$), but not the traffic sign grade in terms of the retroreflective intensity as discussed in~\S\ref{sec:retro_material}. The HIP3930~\cite{3M3930ReflectiveSheeting}, Nikkalite~\cite{Nikkalite92000ReflectiveSheeting}, and DG4090~\cite{3M4000ReflectiveSheeting} are selected to cover the micro prism-based retroreflective materials to cover the ASTM D4956 Grade III-XI and the AASHTO Grade B and D. Due to the nature of the purpose of retroreflective materials, the more expensive materials have the higher retroreflective intensity. For example, the DG4090 is almost 10 times as expensive (724 USD per $\text{m}^2$) as NittoL (80 USD per $\text{m}^2$). We will evaluate the impact of retroreflective material specifications on the attack capability.

        \renewcommand{\arraystretch}{1}
        \setlength{\tabcolsep}{1pt}
        \begin{table}[t!]
            \centering
            \footnotesize
            \caption{Definition of parameters. Parameters are divided into scenario parameters that vary with the environment and attack parameters that an adversary can control.} %
            \vspace{-0.1in}
            \label{Table:thread_model_parameters}
            \begin{tabular}{cc}
                \setlength{\tabcolsep}{1pt}
                \begin{tabular}{c|c}
                    \toprule
                    Scenario  & Parameter          \\
                    Params &  Description              \\ \hline
                    $d_{lon}$ & Distance:Car$\leftrightarrow$ sign                 \\ 
                    $d_{lat}$ &  Distance:Car $\leftrightarrow$ sign               \\ 
                    $h_{s}$ &  Height of sign\\ 
                    $h_{l}$ &  Height of headlight\\ 
                    $L$      & Intensity of ambient light                        \\ 
                    $L_{H}$  & Intensity of headlight                         \\ 
                    \bottomrule
                \end{tabular} 
            \end{tabular}
            \hspace{-3mm}
            \setlength{\tabcolsep}{0.5pt}
            \begin{tabular}{c|c}
                \toprule
                Attack & Parameter          \\ 
                Params & Description\\\hline
                $(x^i_p, y^i_p)$  &  Patch $i$'s Coordinate \\
                $(W^i, H^i)$ & Patch $i$'s width and height\\
                $MPR$ & Maximum ratio of \\
              & patch area to sign bbox \\
                $C^i $ & Patch $i$'s color\\
                $R^i$ & 
                Patch $i$'s retroreflective coef.\\
                \bottomrule
            \end{tabular}
            \vspace{-0.1in}
        \end{table}

\nsubsection{ARP Capability Analysis}\label{sec:patch_feasibility}

We conducted a preliminary capability analysis of the ARP threat model through material characterization of retroreflective patches. Our material analysis reveals that retroreflective patches exhibit substantially different optical properties under varying lighting conditions, enabling high stealth during daytime while producing strong visual perturbations under nighttime headlight illumination. The measured day-night color differences significantly exceed perturbation magnitudes from prior work—for comparison, the L2 norm of the RP$^2$ attack patch area~\cite{adv-patch} in Fig.~\ref{fig:poc} is 90.0. As demonstrated in Fig.~\ref{fig:four_patch_day_night}, this dual-mode material behavior allows ARP attacks to maintain visual inconspicuousness during daytime while causing significant perturbations on stop signs at nighttime. Detailed material characteristics including RGB color measurements and L2 norm analyses across different retroreflective materials and lighting conditions are provided in Table~\ref{table:selected_patch_eval} in Appendix~\ref{sec:detail-retro-analysis}. 
We further investigate the attack performance against TSR models in \S\ref{subsec:eval_dig}.

    \begin{table}[t!]
\centering
\footnotesize
\setlength{\tabcolsep}{0pt}
\renewcommand{\arraystretch}{1}
\caption{Selected retroreflective materials. The 4 patches can cover a wide variety of the retroreflectivity grades defined in ASTM D4956 and AASHTO M268 standards.}

\begin{tabular}{ccccc}
\toprule
 Name &      NittoL~\cite{NittoWideReflect}            &    HIP3930~\cite{3M3930ReflectiveSheeting}         &      Nikkalite~\cite{Nikkalite92000ReflectiveSheeting}         & DG4090~\cite{3M4000ReflectiveSheeting}   \\ \hline
 &   \includegraphics[height=0.8cm]{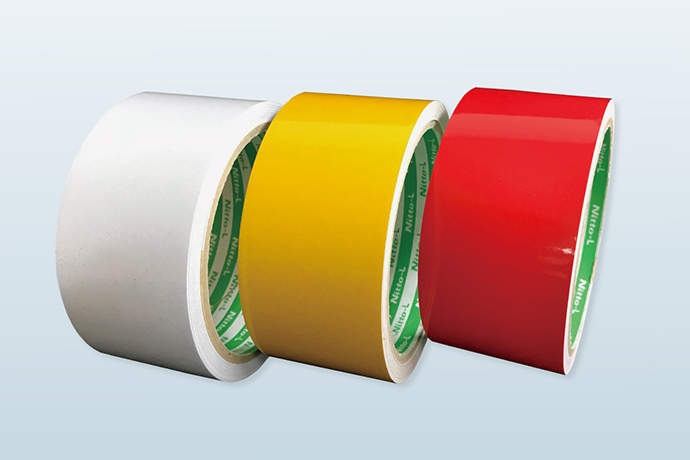}               &    \includegraphics[height=0.8cm]{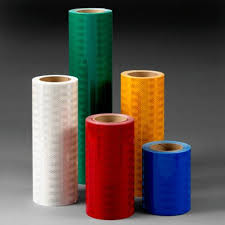}         &    \includegraphics[height=0.8cm]{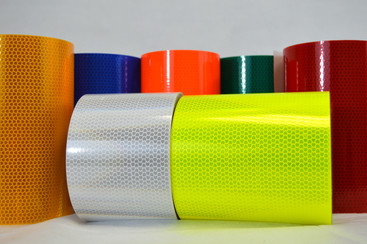}            &         \includegraphics[height=0.8cm]{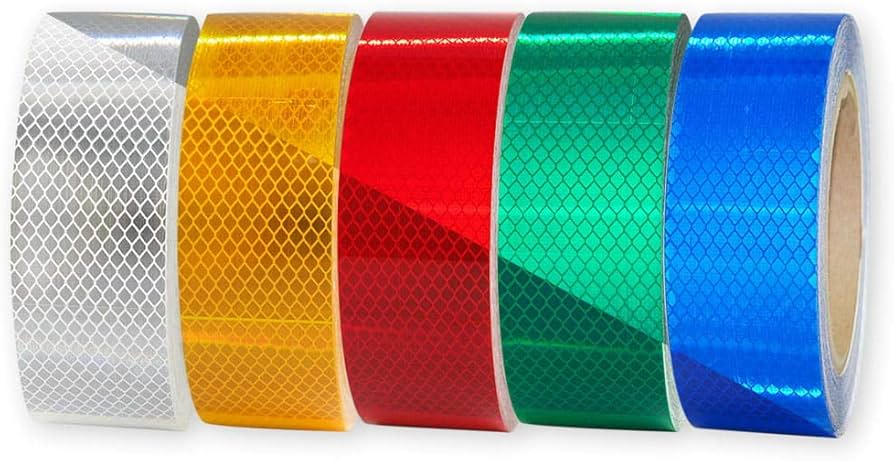}           \\ 
Maker        & Nitto L Material & 3M          & Nikkalite     & 3M                 \\
Brand          & Engineer Grade   & HIP         & Crystal Grade & Diamond Grade \\
Series              & HT               & 3930        & CRG 92000     & 4000               \\
Retroreflector            & Glass Beads      & Micro-Prism & Micro-Prism   & Micro-Prism        \\
Price (USD/m²)    &  \$80           & \$277      & \$479        & \$724             \\\hline
ASTMD4956           & I                & III, IV     & VIII          & XI                \\
AASHTO M268         & N/A              & B           & B             & D                  \\ \toprule
\end{tabular}
\label{table:selected_patch}
\vspace{-0.10in}
\end{table}

%% file: src/032-modeling_attack_optimization.tex
\nsection{Methodology: ARP Attack} \label{sec:methodology}

\begin{figure*}[t]
    \centering
    \includegraphics[width=1.0\linewidth]{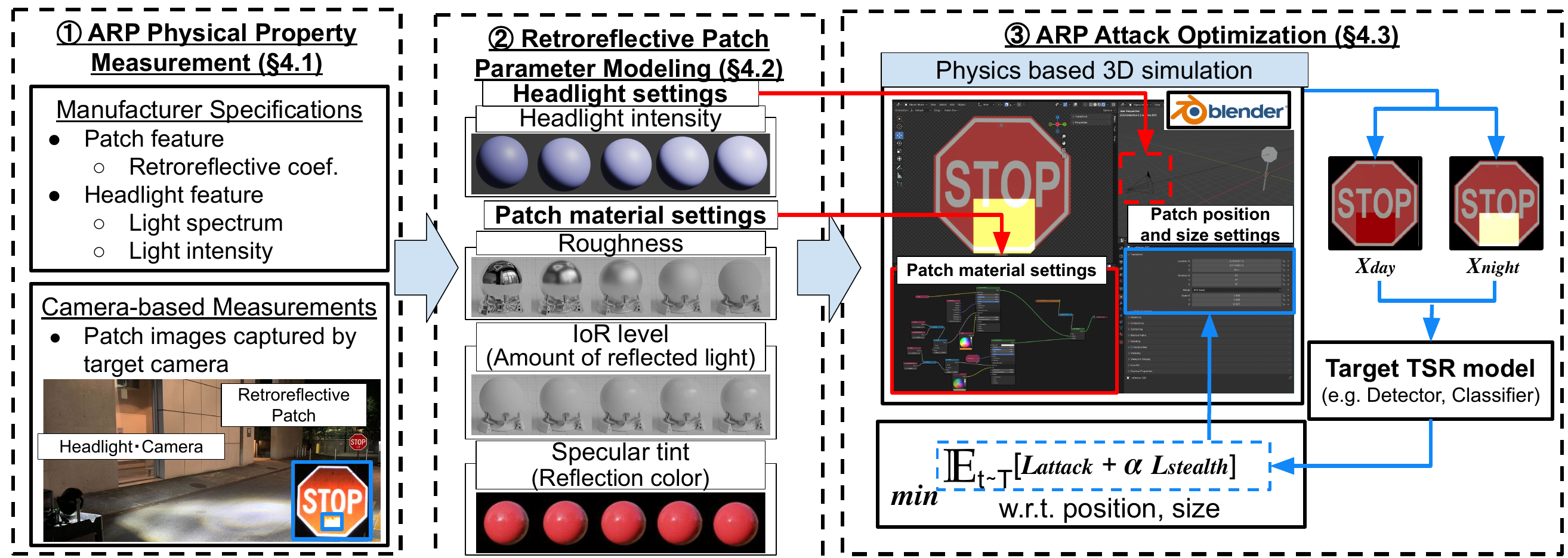}
    \vspace{-0.15in}
    \caption{Overview of ARP Attack Generation. This process involves three main steps: 1. ARP Physical Property Measurement: capturing images with the target camera. 2. Physics-Based Retroreflection Modeling: configuring patch material properties. 3. Day-Night Condition-Based ARP Attack Optimization: optimizing patch position and size}
    \label{fig:attack_overview}
    \vspace{-0.15in}
\end{figure*}

We develop a systematic ARP attack generation pipeline that incorporates the reproduction of retroreflection physical phenomena and retroreflective material properties, enabling accurate simulation and optimization of ARP attacks.
Fig.~\ref{fig:attack_overview} illustrates our three-step approach: (1) we measure the physical properties of actual retroreflective materials (\S\ref{sec:data_collection}) to precisely capture their optical characteristics; (2) we reproduce the retroreflective behavior in Blender (\S\ref{sec:param_estimation}) using a physics-based retroreflection model with parameters from the measurements in (1); and (3) we optimize attack parameters through black-box optimization (\S\ref{sec:attack_generation}) to balance maximizing nighttime attack effectiveness while preserving daylight visual stealth through strategic patch positioning and sizing.
Through this pipeline, we can accurately simulate ARP attacks and efficiently discover attack parameters that are effective in the physical world.

\nsubsection{ARP Physical Property Measurement}\label{sec:data_collection}

To realize Physics-based Retroreflection Modeling, we begin by capturing the real-world optical characteristics of the materials to estimate the parameters to simulate the reflection of retroreflective patches.
We collect the camera images with the camera and headlights used in the target AD vehicle as a first step. We also utilize the specs in the datasheets of the retroreflective patches and the headlight to minimize the effort and to have more reliable data. 
As shown in Fig.~\ref{fig:attack_overview}, we first place the headlight camera at the same height as the target vehicle. We then project the headlight to the retroreflective patches and collect the camera frames. In this experiment, we place the traffic sign with the patch 15 meters away from the camera, corresponding to the ASTM E810 standard test distance for measuring retroreflective properties~\cite{ASTM_E810_20}.

\vspace{-0.05in}
\nsubsection{Retroreflective Patch Parameter Modeling} \label{sec:param_estimation} \label{sec:patch_modeling}
We now detail the core of our physics-based retroreflection model: a novel, physics-grounded simulation approach within Blender designed to accurately reproduce the material's real-world optical characteristics.
A key challenge is that standard 3D shading tools like Blender lack native support for retroreflection, the unique property where light reflects directly back towards its source. To overcome this, we design an approach which involves two main components. First, we design the core retroreflection simulation mechanism using an \textit{imaginary perpendicular reflection plane} to ensure that incoming light returns towards the light source. Second, we establish a parameter optimization process that fit this model using the empirical measurements gathered previously (\S\ref{sec:data_collection}). Specifically, we estimate three key material parameters within Blender's Principled BSDF node~--~Roughness, IoR Level, and Specular Tint~--~which respectively control the diffusion, intensity, and color tint of the simulated reflection. This ensures our simulation faithfully replicates the real-world appearance of the retroreflective patches under investigation.

\nsubsubsection{Imaginary Perpendicular Reflection Plane}

The basic property of the retroreflective light reflection is to reflect the incoming light back to its source direction. 
To simulate the retroreflectivity, we place an imaginary reflection plane perpendicular to the incoming light (see Fig.~\ref{fig:imaginary_plane} for reference).
We can achieve this on Blender by connecting the Incoming property in the Geometry node to the Normal property in the Principled BSDF node. This approach is simple but sufficient to simulate the retroreflectivity on traffic signs.

\nsubsubsection{Reflection Parameters}
We then define the parameters of the imaginary reflection plane. Even if the retroreflective patches try to reflect the incoming light as much as possible, it cannot be perfect and the color of light will be also changed based on the surface film as shown in Fig.~\ref{fig:retroreflective-sheeting-structure}. To simulate the reflection, we calibrate the three parameters: Roughness, IoR Level, and Specular Tint. Roughness can adjust the amount of diffuse reflection. IoR (index of refraction) Level can adjust the intensity of the mirror reflection. Specular Tint can handle the color of reflection lights. We will discuss how to estimate these parameters based on the data we collected.

\noindent\textbf{IoR Level.}
The IoR Level, or reflectance, is defined as the ratio of the flux (or amount of light) actually reflected by a sample surface to the flux that would be reflected by an ideal, perfectly diffuse, completely reflecting standard surface under the same irradiation conditions. 
The relationship between the IoR Level (r) and the retroreflection coefficient (R') can be defined as follows~\cite{field_guide_to_illumination}: 
$
r= \frac{\pi\cdot R'}{\cos\beta \cdot \cos\upsilon} 
$
, where $\beta$ is the entrance angle and $\upsilon$ is the viewing angle. The retroreflection coefficient (R') is always available in the datasheet of the retroreflective materials following the ASTM D4956 or AASHTO standards.

\noindent\textbf{Specular Tint.}
Specular Tint can control the color of reflection light. The reflective color can be calculated by the surface reflectance factor, $R(\lambda)$, the spectral density of the illumination ($S_\lambda(\lambda)$), and the CIE color matching functions ($\bar{x}(\lambda)$, $\bar{y}(\lambda)$, and $\bar{z}(\lambda)$), as expressed in the following equations:

\begin{align}
k &= \frac{100}{\sum_{\lambda=380}^{780} S_\lambda(\lambda) \bar{y}(\lambda)} \\
\begin{pmatrix} X, Y, Z \end{pmatrix} &= k \sum_{\lambda=380}^{780} S_\lambda(\lambda) R(\lambda) \begin{pmatrix} \bar{x}(\lambda) \ \bar{y}(\lambda) \ \bar{z}(\lambda) \end{pmatrix}
\end{align}

Here, the summation is performed over the visible light spectrum, ranging from 380 nm to 780 nm wavelengths.
In this study, we calculate the Specular Tint by using the surface spectral reflectance data in~\cite{reflection_color} and the spectral density of the illumination from \cite{LED_white_spectrum}.

\noindent\textbf{Roughness.}
Roughness controls the minute irregularities on the surface of the retroreflective material and obtain a more realistic retroreflection effect according to the type and characteristics of the retroreflective material. 
As the roughness is not easy to directly measure, we numerically estimate the Roughness, $\alpha$, to minimize the difference between the simulated and observed retroreflective colors as follows:
\begin{align}
    \alpha = \operatorname{argmin}_{\alpha} \left\| C_{\text{real}} - C_{\text{simulated}}(c, r, \alpha) \right\|,
\end{align}
where $C_{\text{real}}$ and $C_{\text{simulated}}$ are the simulated and observed retroreflective colors. $c$ and $r$  are specular tint and IoR Level.

\noindent\textbf{Headlight.}
To simulate the headlights in Blender, the brightness needs to be in watts per square meter~($W/m^2$). Since the typical unit for headlights is lumens (lm), we convert lumens to watts.
At the reference wavelength of 555 nm, where luminous efficacy is 683 lm/W~\cite{SPIELuminousEfficacy}, we calculate the illuminated area $S$ at 15~m as $\pi(15\tan(\alpha))^2$ where $\alpha$ is the spread angle. The Blender irradiance for a headlight of $X$ lumens is thus set to $X/(683S)$, providing accurate headlight modeling.

\nsubsection{
ARP Attack Optimization}\label{sec:attack_generation}

This section describes our approach to solving the remaining two challenges: Stealth-Constrained Patch Optimization and Day-Night Opposing-Objective Optimization.
We first detail the mechanism for enforcing the stealth constraint, and then formulate the multi-objective optimization problem.

\noindent\textbf{Stealth-Constrained Patch Optimization.} We enforce the color-matching constraint directly within our Blender pipeline. Patches must precisely match the background color of their placement location (e.g., white patches for white areas of a STOP sign, red patches for red areas). We achieve this automatically using Blender's shader node system; specifically, a Mix Shader allows us to dynamically select the appropriate retroreflective material based on the sign's underlying color at any given position. This ensures any patch generated during optimization inherently satisfies the stealth constraint.
With this constraint mechanism in place, we designed the optimization-based ARP attack generation.

\noindent\textbf{Day-Night Opposing-Objective Optimization.} With this constraint mechanism in place, and based on the collected data (\S\ref{sec:data_collection}) and the retroreflective patch modeling~(\S\ref{sec:patch_modeling}), we designed the optimization-based ARP attack generation. 

First, let $X_{day}$ and $X_{night}$ represent the simulated images of the traffic sign with a candidate ARP applied, rendered under daytime (ambient light) and nighttime (vehicle headlights active) conditions, respectively. These images are generated by our Blender-based simulation functions, \texttt{SimulateARPday} and \texttt{SimulateARPnight}, which take the patch parameters $(x_p, y_p, W, H)$ as inputs.
We then formulate an optimization problem to find the best patch parameters by balancing attack effectiveness and daytime stealth, expressed as:

\begin{align}
\min_{x_p, y_p, W, H} &\mathbb{E}_{t \sim \text{T}} [L_{attack}(X_{night}) + \alpha L_{stealth}(X_{day})]\label{eq:optimization}\\
\text{s.t.} \quad & X_{day} = \text{SimulateARP}_{day}(x_p, y_p, W, H) \nonumber\\
& X_{night} = \text{SimulateARP}_{night}(x_p, y_p, W, H) \nonumber\\
& W \times H \leq MPR \cdot W_{bbox} \cdot H_{bbox}\nonumber,
\end{align}

where, $(x_p,y_p)$ represents the patch position, and $(W,H)$ is the patch size, which are our primary optimization parameters. We define $MPR$ as a hyperparameter that constrains the total patch area relative to the sign's bounding box area $(W_{bbox} \times H_{bbox})$ to incorporate a unified index. For instance, for a speed limit sign measuring 24 inches × 30 inches (720 square inches), a $MPR$ of 0.0625 corresponds to 45 square inches (approximately 0.029 square meters). This ratio-based constraint ensures that our patch size scales appropriately with different sign dimensions.

The objective function in Eq.~\ref{eq:optimization}, balances two components with the hyperparameter $\alpha$: the nighttime attack loss ($\mathcal{L}_{\text{attack}}$) and the daytime stealth loss ($\mathcal{L}_{\text{stealth}}$). The attack loss is defined as $\mathcal{L}_{\text{attack}} = L(X_{\text{night}})$, where $L(X)$ is the model's confidence score for the true sign class. Minimizing this term encourages misclassification or misdetection at night. Conversely, the stealth loss is defined as $\mathcal{L}_{\text{stealth}} = -L(X_{\text{day}})$, which maximizes the true class confidence during the day to ensure the patch does not disrupt normal TSR operation. Through empirical evaluation, we found $\alpha = 1$ works best.

To ensure robustness across various environmental conditions, we apply the Expectation over Transformation (EoT) technique~\cite{athalye2018synthesizing}, applying random transformations~($T$) on lighting and camera angles for the simulated images within Blender. 
Since our primary contribution focuses on the attack pipeline design rather than optimization methodology itself, and given the non-differentiable nature of our Blender-based simulation, we utilize a well-established black-box optimization method, the Tree-structured Parzen Estimator algorithm~\cite{TPE-algorithm} implemented in the Optuna~\cite{Optuna} framework. This approach allows us to efficiently explore the parameter space and converge on an effective attack configuration.
Through this optimization process, we aim to generate ARPs that remain inconspicuous during daylight hours while consistently degrading the performance of the target TSR system in low-light conditions. 

%% file: src/04-Evaluation1.tex
\nsection{Evaluation of ARP Attack}\label{sec:evaluation}

We systematically evaluate the ARP Attack to validate its core design goals: effectiveness and stealthiness. First, we validate attack effectiveness through digital environment evaluations, establishing pipeline validity via baseline comparisons and investigating performance across four material types (\S\ref{subsec:eval_dig}), then physically implement the ARP attack to evaluate robustness against real-world variables; distance, angle, and ambient light (\S\ref{sec:eval_real}). Finally, we assess attack stealthiness from both objective and subjective perspectives (\S\ref{sec:eval_stealth}).

\nsubsection{Foundational Validation in Digital Space}\label{subsec:eval_dig}
We first systematically evaluate the effectiveness of the ARP attack in the digital space using Blender~\cite{Blender}, a state-of-the-art 3D shading simulator, 

 We then evaluate attack transferability across different model architectures and training datasets.

\nsubsubsection{Experimental Setup}

     We setup the threat model as described in~\S\ref{sec:threat} using Blender.
    We place U.S.-style traffic signs -- a stop sign (STOP) and a 65 mph speed limit sign (SL65) -- at a longitudinal distance ($d_{lon}$) of 15 m from the target camera and the headlight with no lateral offset ($d_{lat}$ = 0 m).
    The height of the traffic signs ($h_s$) is set at 1.5 m, following the previous research~\cite{jia2022foolingeyesautonomousvehicles}.
    The vehicle headlight is positioned at a height ($h_l$) of 0.75 m, estimated from the technical specifications of the Tesla Model S~\cite{Tesla2020Manual}. 
    The headlight parameters are configured based on industry standards, with a spread angle of 5 radians~\cite{De_2019_vehicleheadlight} and luminous flux of 3,400 lumens~\cite{BAOLICY2024}. These specifications are implemented in Blender following the methodology detailed in~\S\ref{sec:methodology}, to accurately represent real-world lighting conditions and attack scenarios.
    
    For simulation in Blender, we precisely replicate the characteristics of the FLIR BlackflyS Machine Vision Camera (BFS-PGE-16S2C-CS), a camera commonly adopted in the AD vehicles integrated with Autoware~\cite{autoware}. 
    The virtual camera configuration incorporates a 1/1.8 inch sensor (7.2 mm × 5.4 mm), 12 mm focal length, and native resolution of 1440 × 1080 pixels, ensuring our simulated image capture accurately reflects the field of view and imaging characteristics of actual AD vision systems.

\textit{1)~Target Traffic Sign Recognition Models:}
    For the single-stage architecture, we employ YOLOv5~\cite{YOLO} and Faster R-CNN~\cite{Faster-RCNN}, as they are two major architectures in object detection tasks.
        These models are trained on the ARTS~\cite{ARTS_dataset} and Mapillary~\cite{mapillary_dataset} datasets which are the widely used publicly available US-style traffic sign datasets for object detection and tried our best to train high-performance TSR models.
    As the Mapillary dataset contains traffic signs from various countries, we extract and use only U.S.-style speed limit and stop signs to train the model for the evaluation.
    For all object detection models, the confidence threshold was set to 0.3 by following previous work~\cite{jia2022foolingeyesautonomousvehicles}.
    For the two-stage architectures, we crop the Region of Interest (ROI) containing the target signs with the ground truth and evaluate their classification directly by the second-stage classifier as our analysis focuses on the classification process.
    For our evaluation, we adopt SimpleCNN architecture used in previous works~\cite{sato2024invisible}.
    We train this model on three standard datasets: GTSRB~\cite{GTSRB}, LISA~\cite{LISA_dataset}, and ARTS~\cite{ARTS_dataset}.
    Table~\ref{table:model-performance} in Appendix~\ref{sec:appendix_benign-model-performance} lists the benign performances of the targeted TSR models.

\textit{2)~ Evaluation Metrics:} 
     We use the Attack Success Rate (ASR) as a main evaluation metric, following the previous studies~\cite{DBLP:journals/corr/abs-1807-07769,SLAP,sato2024invisible}. 
    The ASR is calculated as the proportion of instances where the target traffic sign is misclassified or fails to be detected by the TSR system. 

\nsubsubsection{Attack Effectiveness Evaluation}
We first evaluate the attack effectiveness to see if our method can actually find effective ARP attacks and the necessity of physics based reflection color modeling.
. We then analyze the impact of patch size and the number of patches.
    
    \noindent\textbf{Attack Effective over Baseline.}
    To highlight the effectiveness of our attack generation methodology, we compare it against a random baseline attack, which places a randomly-sized square patch at a random location on the sign. We generated 100 instances for two MPR values (0.1875 and 0.25) using the highest retroreflective performance DG4090 material.  Table~\ref{tab:attack_effectivness} show that ARP attack achieves significantly higher ASR than the random baseline across all tested models, confirming that our optimization process is crucial for finding effective attacks.

    \begin{table}
    \centering
    \footnotesize
    \caption{Attack effectiveness of single- and two-stage
    TSR systems with ASR. Bold numbers indicate the higher ASR between the random baseline attack and ARP Attack.} 

    \label{tab:attack_effectivness}

    \begin{tabular}{cccccccc}
    \toprule
     &                         &       Attack Type              & \multicolumn{2}{c}{Random Attack} &  & \multicolumn{2}{c}{ARP Attack} \\ \cline{4-5} \cline{7-8} 
     &                               &              $MPR$               & 0.1875             & 0.25            &  & 0.1875             & 0.25            \\ \hline\hline
    \multirow{6}{*}{\rotatebox{90}{STOP}}   & \multirow{4}{*}{Single-Stage} & Faster R-CNN (ARTS) & 1\% & 3\% &  & \textbf{100\%} & \textbf{90\%} \\
    &                            & Faster R-CNN (Mapillary) & 1\%              & 0\%            &  & \textbf{5\%  }        & 0\%         \\
     &                            & YOLOv5 (ARTS)          & 13\%                & 8\%            &  & \textbf{100\%}          & \textbf{90\%}           \\

     &                            & YOLOv5 (COCO)          & 0\%                & 0\%            &  & 0\%          & \textbf{68\%}           \\ \cline{2-8} 
     & \multirow{2}{*}{Two-Stage}& SimpleCNN (ARTS)              & 8\%             & 5\%            &  & \textbf{100\%}          & \textbf{99\%}          \\
     &                            & SimpleCNN (GTSRB)             & 24\%             & 19\%            &  & \textbf{100\%}          & \textbf{100\%}          \\ \hline\hline
    \multirow{4}{*}{\rotatebox{90}{SL65}} & \multirow{2}{*}{Single-Stage} & Faster R-CNN (ARTS) & 0\% & 0\% &  & \textbf{100\%} & \textbf{100\%} \\
     &                            & Faster R-CNN (Mapillary) & 5\%                & 2\%            &  & \textbf{85\%}          & \textbf{80\%}           \\ \cline{2-8} 
     & \multirow{2}{*}{Two-Stage} & SimpleCNN (ARTS)              & 6\%             & 2\%            &  & \textbf{100\%}          & \textbf{100\%}         \\
     &                            & SimpleCNN (LISA)              & 15\%             & 23\%            &  & \textbf{100\%}          & \textbf{100\%}         \\ \bottomrule
    \end{tabular}
\vspace{-0.1in}
    \end{table}

       \noindent\textbf{Necessity of our Physics-based Retroreflective Simulation}
       To demonstrate the necessity of our physics-based simulation, we compared its effectiveness against a naive "white assumption" baseline used in prior work~\cite{chen2024reflective}, which assumes that reflections are uniformly white. In this ablation study, the baseline optimizes patch placement assuming uniform white reflection, while our method directly optimizes using our physics-based model. As shown in Tab.~\ref{tab:ablation_study}, our physics-based modeling significantly improves attack performance, especially for lower-grade materials (NittoL, HIP3930), where the ASR increased by 20-50 percentage points. This result demonstrates two key points: the naive white assumption severely underestimates the attack potential of realistic materials, and our physics-based approach enables effective attacks even with more economical materials.

    \begin{table}
            \centering
            \footnotesize
            \caption{Comparison White Assumption with Physics-Based Modeling with ASR. Bold numbers indicate higher ASR.}

            \label{tab:ablation_study}
            \begin{tabular}{cccccccc}
            \toprule
             &    Patch                     &       Attack Type              & \multicolumn{2}{c}{White Assumption} &  & \multicolumn{2}{c}{Physics-Based (Ours)} \\ \cline{4-5} \cline{7-8} 
             &       Material                         &              $MPR$               & 0.1875             & 0.25            &  & 0.1875             & 0.25            \\ \hline\hline
              & NittoL & & 25\% & 40\% &  & \textbf{70}\% & \textbf{90}\%\\
            &  HIP3930 & & 40\% & 80\% &  & \textbf{60}\% & \textbf{90}\%\\
            & Nikkalite & & 35\% & 65\% &  & \textbf{50}\% & \textbf{90}\% \\
            & DG4090 & & 75\% & \textbf{100}\% &  & \textbf{90}\%& 95\% \\ \bottomrule
            \end{tabular}
        \vspace{-0.1in}
    \end{table}

\noindent\textbf{Impact of Patch Size and Patch Material.}
    We then analyze the impact of patch size and patch retroreflective material.
    We evaluate using YOLOv5 for single-stage architecture and SimpleCNN for two-stage architecture trained on the ARTS dataset, as it provides comprehensive coverage of U.S. traffic signs.
    We evaluate ASR for each combination of four $MPR$ values (0.0625, 0.125, 0.1875, 0.25) and four retroreflective materials listed in Table~\ref{table:selected_patch}.
    More details are in Appendix \ref{Append:Attackeffect}.

    These results indicate effective attack configurations: For STOP sign, DG4090 material with $MPR$ of 0.1875 achieves optimal performance (100\% ASR) against single-stage architectures, while 0.125 $MPR$ is required for maximum effectiveness against two-stage architectures. SL65 sign attacks can be successfully executed using the more economical NittoL with the minimal 0.0625 $MPR$ for both architectures.

\noindent\textbf{Attack Impact of Multiple Patches.}
    We investigate the impact of 
    From our results, we find that using a single, larger patch is more effective than multiple smaller patches. 
    More details are in Appendix \ref{Append:Attackeffect}.

    \noindent\textbf{Attack Transferability Evaluation.}We also evaluate the transferability of ARP attacks to other TSR models with different architectures and datasets.
    We follow the same experimental setup as in \S\ref{subsec:eval_dig}. 
The ARP attack transferability to different architectures shows 100\% ASR for SL65 signs but not for STOP signs. The attack transferability varies from 55\% to 90\% ASR depending on the source and target dataset pair. For the different datasets, the attack transferability may be affected by whether the attack compromised class exists on the destination dataset or not. See Appendix ~\ref{appendix_attack_trans} for details.

%% file: src/05-Evaluation2.1.tex
\nsubsection{Physical-World Robustness Evaluation}
    \label{sec:eval_real}

    We evaluate ARP attack robustness across various environmental conditions including vehicle positions, camera heights, sign elevations, and ambient lighting to verify that digitally generated attacks maintain effectiveness in physical deployment. Interested readers may refer to Appendix~\ref{sec:appendix-multiple-vehicle} for robustness evaluation to multiple vehicles.

\begin{figure}[t!]
    \centering
    \includegraphics[width=0.95\linewidth]{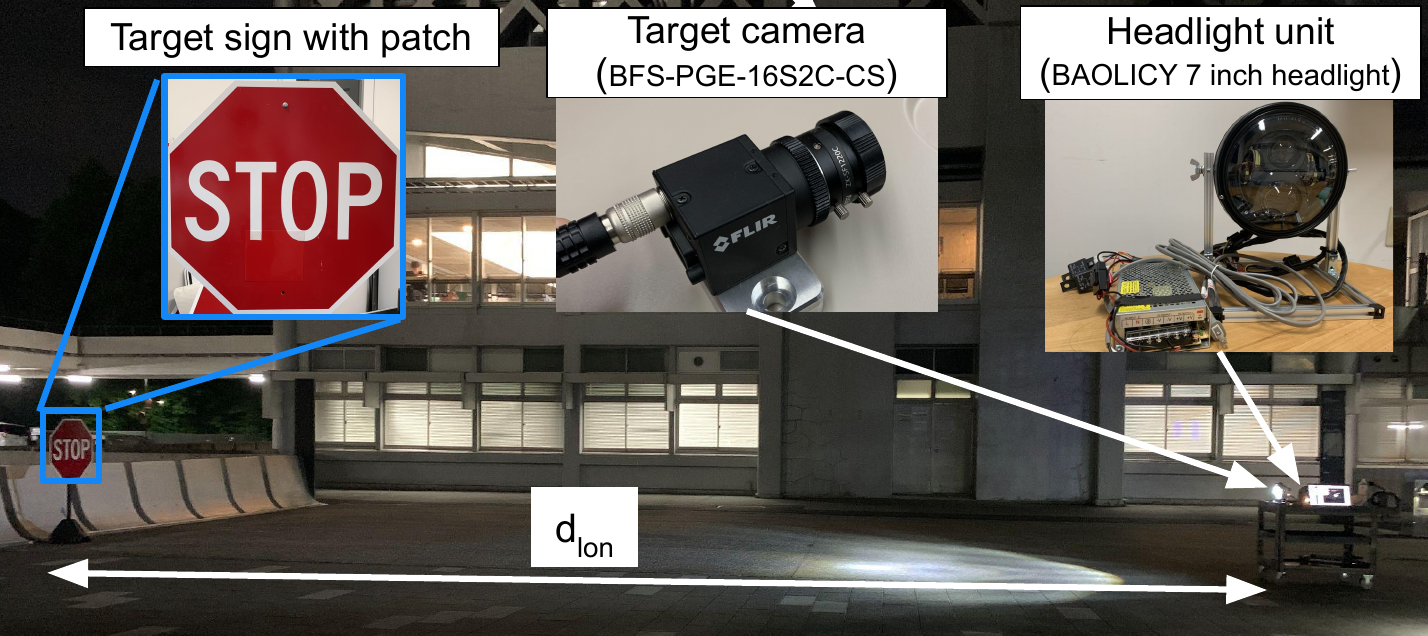}
    
    \caption{Experimental setup of physical world experiment. The headlight and camera are set on top of the carriage.}
    \vspace{-0.1in}
    \label{fig:Experimental_setup}

\end{figure}

\nsubsubsection{Experimental Setup}
    We set the same conditions as the simulation experiment conducted in \S\ref{subsec:eval_dig}, as shown in~Fig.~\ref{fig:Experimental_setup}. For outdoor experiments, we targeted YOLOv5 as a single-stage architecture model and SimpleCNN as a two-stage architecture model, both trained on the ARTS dataset. To capture images, we used a FLIR Machine Vision Camera BlackflyS (BFS-PGE-16S2C-CS)\cite{FLIRBlackflySGigE}, which is referenced in Autoware~\cite{autoware}, mounted on the target AD vehicle. The vehicle was equipped with headlights having a brightness of 3400 lumens \cite{De_2019_vehicleheadlight}.

    We optimized adversarial patches with configurations specific to each target sign and model architecture, based on the analysis in~\S\ref{subsec:eval_dig}. In attacking STOP signs, DG4090 reflective material was used for both architectures, with $MPR$ of 0.1875 and 0.125 for single-stage and two-stage architectures, respectively. The SL65 sign attacks utilized NittoL material across both architectures with a $MPR$ of 0.0625. We evaluated the attack effectiveness using ASR as described in~\S\ref{subsec:eval_dig}. For each configuration, we captured 100 images and calculated the ASR.

\begin{figure}[t!]
    \centering
    \includegraphics[width=\linewidth]{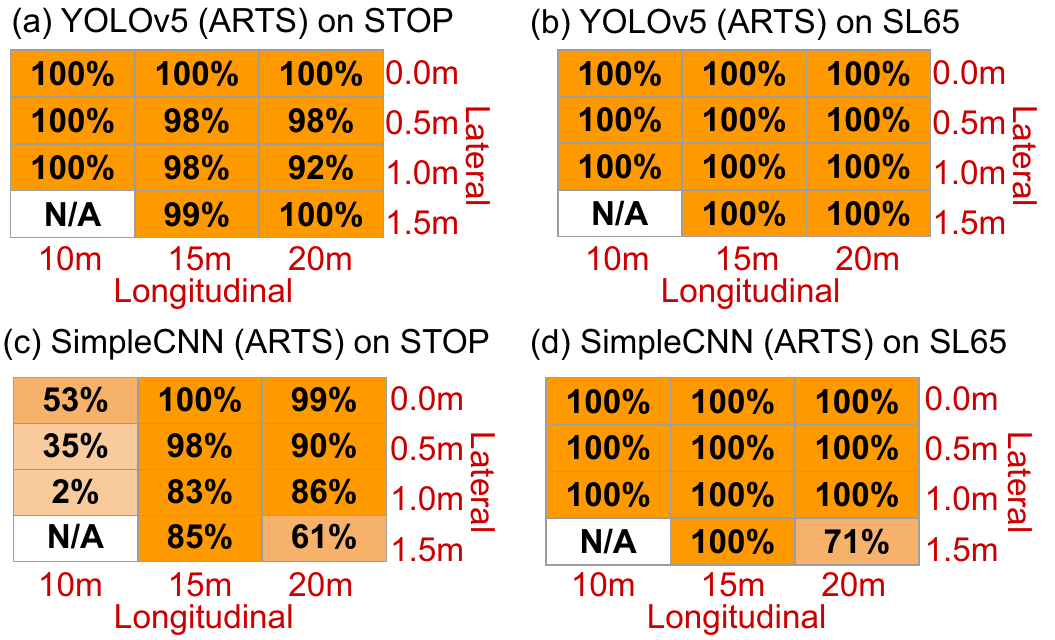}

    \caption{ASR for single- and two-stage TSR in different camera positions. N/A: traffic sign is not visible in the images. }
        \label{fig:phys_eval}
        \vspace{-0.2in}
\end{figure}

\nsubsubsection{Results}
We will demonstrate the robustness of ARP attacks under varying environmental conditions, including different vehicle positions and ambient lighting levels.
\noindent\textbf{\\Robustness of Different Target Vehicle Positions.}
    Fig.~\ref{fig:phys_eval} shows the attack performance against single- and two-stage architectures. All attacks were optimized at a fixed distance ($d_{lon} = 15~\text{m}, d_{lat} = 0~\text{m}$). 
    The ARP attacks maintained high effectiveness across most tested distances.
    This is due to the retroreflective material's property of returning light toward its source.
    However, attacks on STOP signs showed reduced effectiveness against two-stage architectures at close distances, as steep angles diminished the light reaching the patch, weakening its attack effect. 
    From a practical perspective, this performance degradation at close distances does not significantly impact the attack's real-world effectiveness. By the time a vehicle reaches these close distances (under 15~m), it has already needed to make critical decisions about stopping. We further evaluate driving experiments in~\S\ref{sec:driving}.

\noindent\textbf{Robustness to Different Ambient Light Conditions.}
We investigate how ambient light conditions affect ARP attack effectiveness. We conducted physical experiments on both STOP and SL65 signs during the transition from daylight to darkness. We measured ASR at 5-minute intervals: from 16:40 to 19:00 (sunset at 18:00) for STOP signs and from 16:40 to 18:40 (sunset at 17:50) for SL65 signs. While we conducted experiments on different days, we maintained consistent parameters by conducting all experiments under clear weather conditions in the same urban environment. Throughout our evaluations, we used a YOLOv5 model trained on the ARTS dataset as our single-stage TSR target.

\begin{figure}
        \centering
        \includegraphics[width=  1\linewidth]{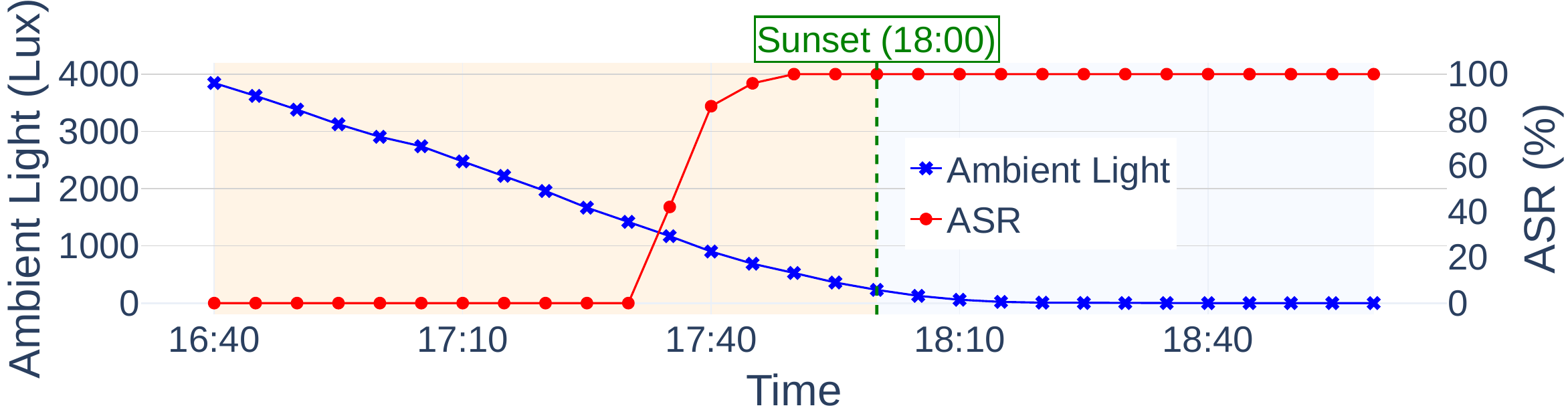}
        \vspace{-0.25in}
        \caption{Ambient light intensity (lux) and ASR over time. Sunset is 18:00. ARP attack achieves high attack performance ($\geq$85\% ASR) at $\leq$900 lux.}
        \label{fig:light-level-asr-stop}
        \vspace{-0.15in}
    \end{figure}

\begin{figure}
        \centering
        \includegraphics[width=  1\linewidth]{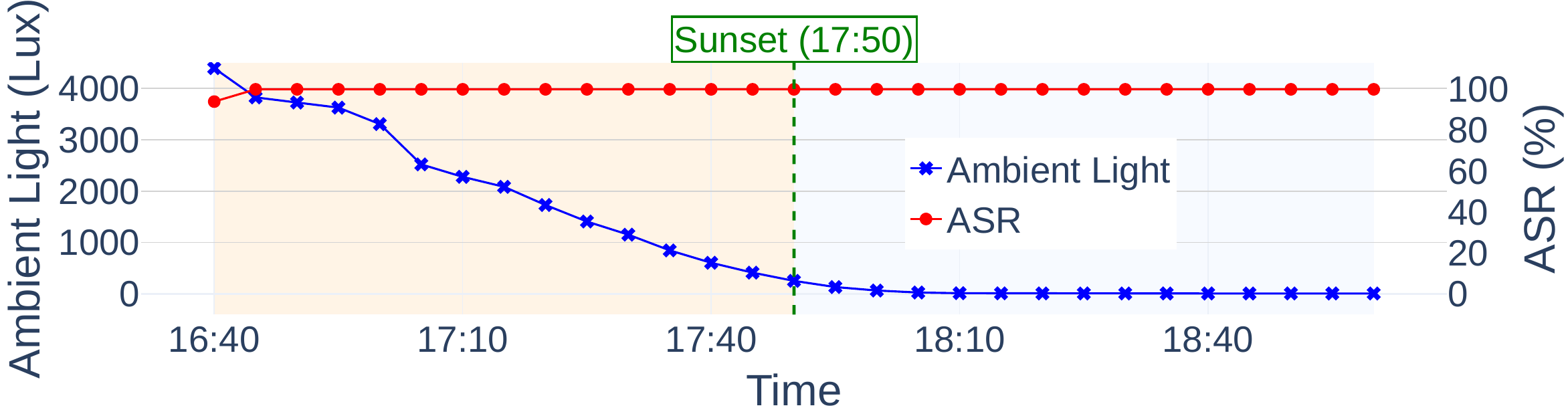}
        \vspace{-0.25in}
        \caption{Ambient light intensity (lux) and ASR over time. Sunset is 17:50. ARP attack achieves high attack performance ($\geq$90\% ASR) at $\leq$4000 lux.}
        \label{fig:light-level-asr-sl65}

    \end{figure}

    As shown in Fig.~\ref{fig:light-level-asr-stop} and Fig.~\ref{fig:light-level-asr-sl65}, our results show different robustness in ambient light conditions depending on the sign type. 
    The attacks against STOP signs only became effective when ambient light dropped below 900 lux—approximately 20 minutes before sunset in our testing environment. As light levels decreased further, attack effectiveness increased dramatically, with ASR rising from 41.5\% at 1170 lux to 85.9\% at 900 lux, eventually reaching 100\% at 530 lux (about 10 minutes before sunset). In contrast, SL65 sign attacks maintained over 90\% effectiveness even in early evening conditions with ambient light up to 4000 lux, demonstrating efficacy across a much wider range of lighting conditions.
ARP attack demonstrates higher robustness to ambient light than prior light or laser projection attacks~\cite{sato2024invisible}. 
This robustness stems from the physical properties of retroreflection, which directively reflect back toward the source, maintaining high attack effectiveness even in brighter surroundings.

As demonstrated, the ARP attack shows high robustness against real-world variables such as vehicle distance and ambient light. To further validate this robustness comprehensively, we also conducted evaluations under more diverse conditions, including variations in sign height, camera mounting position, and even traffic scenarios with multiple vehicles present. The attack effectiveness was maintained across these conditions, with detailed results available in Appendix \ref{appendix_robust_sign_camera} and \ref{sec:appendix-multiple-vehicle}.

%% file: src/05-Evaluation2.2-Stealthiness.tex
\nsubsection{Stealthiness Evaluation} \label{sec:eval_stealth}

Finally, we evaluate the stealthiness of the ARP attacks through both {\it objective} perceptual similarity metrics and {\it subjective} human perception assessments, expecting our approach to demonstrate superior stealthiness compared to prior patch attacks under both inactive (daytime) and active (nighttime) conditions. 
For objective evaluation, we employ widely-recognized perceptual similarity metrics SSIM and LPIPS to quantitatively measure visual discrepancies between benign and adversarial traffic signs. 
As detailed in Appendix~\ref{sec:setalth_perceptual}, our ARP attack achieves significantly higher stealthiness scores than prior patch attacks, with improvements of $\geq$0.1 and 0.15 points for SSIM and LPIPS respectively, demonstrating substantially reduced visual artifacts.

To complement the objective analysis, we conducted a comprehensive user study with 50 participants to assess subjective human perception of naturalness across various viewing scenarios including different distances, times of day, and viewing perspectives (driver view and pedestrian view). 
Due to space constraints, the detailed experimental methodology and user study design are provided in Appendix~\ref{Append:stealth-user} and Appendix~\ref{Append:stealth-user-detail}, respectively. 
The results demonstrate that our ARP attacks consistently received high naturalness ratings (mostly 1.71--2.29 on a 5-point scale) that closely match those of unmodified signs (1.81), while prior attack methods received significantly higher unnaturalness ratings (3.69--4.19).
Notably, our approach maintained high stealthiness even during nighttime scenarios when the retroreflective attack was actively functioning, demonstrating the effectiveness of leveraging natural physical phenomena rather than artificial patterns.

%% file: src/06-DrivingEvaluation.tex
\nsection{Driving Evaluation on Real Car}\label{sec:driving}

\begin{figure}[t!]
     \centering
     \includegraphics[width=\linewidth]{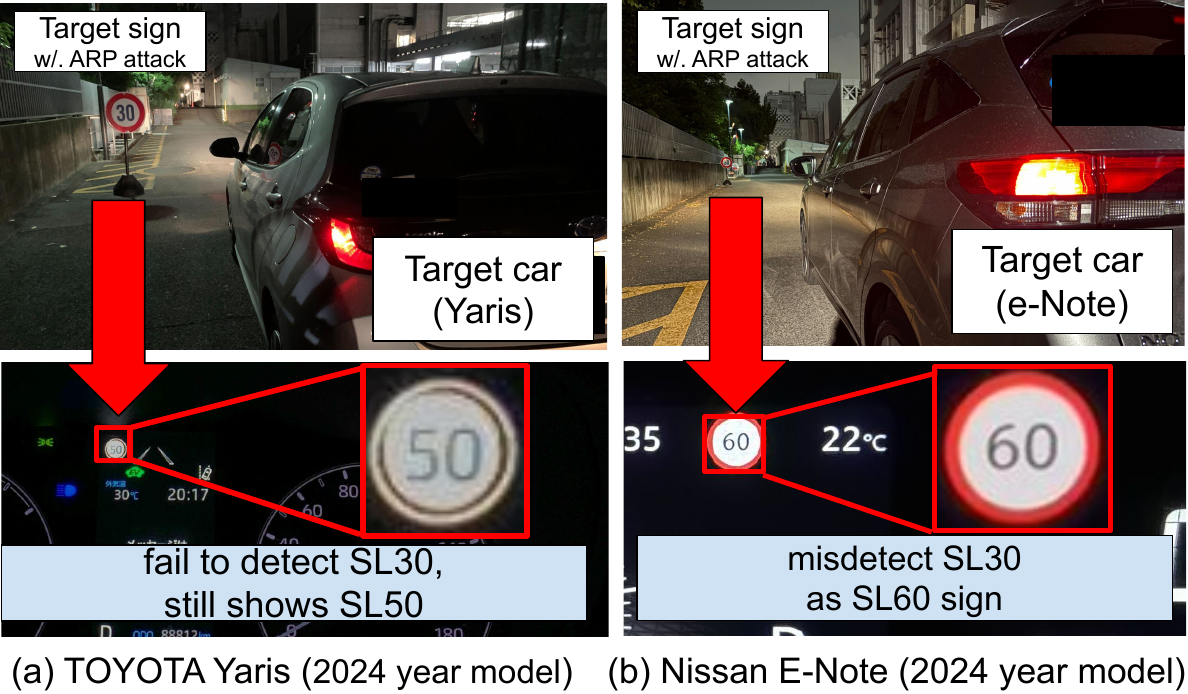}
     \vspace{-0.2in}
     \caption{Results of the ARP attack on commercial TSRs. Our attack causes the TSRs to misdetect the 30 km/h sign, causing  (a) the Yaris to fail to update its 50 km/h display, and (b) the e-Note to misclassify the sign as 60 km/h}
     \label{fig:commercial-TSR}

\end{figure}

    To evaluate the real-world impact of ARP attacks on a real TSR system, we conducted comprehensive experiments under realistic driving conditions using both custom-trained models and a commercial TSR system.
    
    \nsubsection{Evaluation on Research-Grade TSR Model}
    We first evaluated our attack against custom-trained TSR models in controlled driving scenarios to systematically assess its effectiveness across varying distances.

    \noindent\textbf{Experimental Setup.}
    We employed the same target models and camera configurations as described in \S~\ref{sec:evaluation}. We used the optimal attack configurations identified in~\S\ref{subsec:eval_dig}: DG4090 material with $MPR$ of 0.1875 and 0.125 for STOP signs (single-stage and two-stage architectures respectively), and NittoL material with $MPR$ of 0.0625 for the SL65 sign.
    Fig.~\ref{fig:driving} in Appendix\ref{sec:driving-setup} illustrates our experimental setup. We mounted a camera at the center of a Toyota Yaris~\cite{ToyotaYaris2024} dashboard, positioned behind the windshield with its optical axis aligned with the vehicle's longitudinal axis. We recorded videos while driving the vehicle toward the traffic sign. Starting from 50 meters away, we maintained a constant speed of 5, 15, 25 km/h (approximately 3,9,15 mph) during the approach to ensure safe and controlled vehicle operation.
    To quantify attack effectiveness, we calculated the ASR across different distances. We divided the approach path into 5-meter intervals from 50~m to 15~m (the closest distance at which the sign remains fully visible in the camera frame). The ASR for each interval represents the percentage of frames where the attack successfully caused misclassification or misdetection. We performed five driving trials and aggregated the results across all trials to compute the 
    ASR for each distance interval.

\noindent\textbf{Results.}
The real-world driving tests showed the effectiveness of ARP attacks across the range of speeds, distances, and signs tested.
At ranges beyond 35 m, our attacks achieved consistently high ASRs ($>$93\%) against both STOP and SL65 signs, regardless of vehicle speed (5-25 km/h). SL65 sign attacks proved particularly robust, maintaining 95-100\% ASR across all tested distances (15-50 m) and speeds. STOP sign attacks, while highly effective at distances beyond 35 m,  maintained substantial effectiveness (50-75\% ASR) at closer ranges ($<$ 30 m), with variations depending on the speed.

    While SL65 sign attacks demonstrated consistent effectiveness across all conditions, the attack effectiveness against STOP signs presents another serious vulnerability, with high ASR observed at distances beyond 35 m. While our experiments were limited to 25 km/h for safety reasons, the consistent attack effectiveness across our tested parameters suggests potential vulnerability at the distance where critical stopping decisions must be made. The attack's effectiveness at ranges beyond 35~m is particularly concerning, as this distance corresponds to the typical braking distance for vehicles at urban speeds~\cite{DOT_FHWA_2009}. This means our attack can compromise the TSR system precisely within the critical decision-making window, posing significant risks in areas like residential neighborhoods.

\newcommand{\stackth}[1]{\begin{tabular}{@{}c@{}}#1\end{tabular}}

\begin{table}[t]
    \footnotesize
    \centering
    \setlength{\tabcolsep}{1pt} 
    \renewcommand{\arraystretch}{0.9}
    
    \caption{ASR of the ARP Attack in driving scenarios with at different speeds. Bold numbers indicate ASR above 80\%.}
    \label{tab:attack-model-performance-final-percent}

    \begin{tabular}{@{}l l c ccccccc@{}} 
    \toprule
    \multirow{2}{*}{Sign} & \multirow{2}{*}{Model} & \multirow{2}{*}{\stackth{Speed\\(km/h)}} & \multicolumn{7}{c}{ASR at Distance (m)} \\
    \cmidrule(l){4-10}
    & & & \stackth{15-20} & \stackth{20-25} & \stackth{25-30} & \stackth{30-35} & \stackth{35-40} & \stackth{40-45} & \stackth{45-50} \\
    \midrule
    \multirow{6}{*}{STOP} & \multirow{3}{*}{\scriptsize{\stackth{YOLO\\v5}}} & 5 & 61.7\% & 72.2\% & 76.4\% & \textbf{99.8\%} & \textbf{94.7\%} & \textbf{97.3\%} & \textbf{100\%}\\
    & & 15 & 73.5\% & 51.4\% & 64.5\% & 73.0\% & \textbf{99.8\%} & \textbf{100\%} & \textbf{100\%} \\
    & & 25 & 44.5\% & \textbf{91.3\%} & \textbf{100\%} & \textbf{100\%} & \textbf{100\%} & \textbf{100\%} & \textbf{100\%} \\
    \cmidrule(l){2-10}
    & \multirow{3}{*}{\stackth{\scriptsize{Simple}\\\scriptsize{CNN}}} & 5 & 55.6\% & 60.9\% & 70.1\% & 67.1\% & \textbf{93.4\%} & \textbf{98.6\%} & \textbf{94.9\%} \\
    & & 15 & 78.2\% & 75.9\% & \textbf{98.3\%} & \textbf{100\%} & \textbf{98.1\%} & \textbf{96.4\%} & \textbf{94.7\%} \\
    & & 25 & 28.0\% & 60.8\% & \textbf{95.6\%} & \textbf{90.2\%} & \textbf{93.6\%} & \textbf{87.2\%} & \textbf{89.0\%} \\
    \midrule
    \multirow{6}{*}{SL65} & \multirow{3}{*}{\scriptsize{\stackth{YOLO\\v5}}} & 5 & \textbf{100\%} & \textbf{100\%} & \textbf{100\%} & \textbf{100\%} & \textbf{100\%} & \textbf{100\%} & \textbf{100\%} \\
    & & 15 & \textbf{100\%} & \textbf{100\%} & \textbf{100\%} & \textbf{100\%} & \textbf{100\%} & \textbf{100\%} & \textbf{100\%} \\
    & & 25 & \textbf{100\%} & \textbf{100\%} & \textbf{100\%} & \textbf{100\%} & \textbf{100\%} & \textbf{100\%} & \textbf{100\%} \\
    \cmidrule(l){2-10}
    & \multirow{3}{*}{\stackth{\scriptsize{Simple}\\\scriptsize{CNN}}} & 5 & \textbf{99.7\%} & \textbf{96.2\%} & \textbf{98.9\%} & \textbf{99.4\%} & \textbf{99.5\%} & \textbf{100\%} & \textbf{100\%} \\
    & & 15 & \textbf{99.6\%} & \textbf{100\%} & \textbf{99.8\%} & \textbf{100\%} & \textbf{100\%} & \textbf{100\%} & \textbf{100\%} \\
    & & 25 & \textbf{100\%} & \textbf{99.7\%} & \textbf{100\%} & \textbf{100\%} & \textbf{98.6\%} & \textbf{99.6\%} & \textbf{100\%} \\
    \bottomrule
    \end{tabular}
\vspace{-0.1in}
\end{table}

\nsubsection{Evaluation on the Production TSR System}
    To demonstrate real-world applicability, we further evaluated our attack against production TSR system installed in commercial vehicles.

\noindent\textbf{Experimental Setup.}
    We tested our ARP attack against production TSR systems installed in a Toyota Yaris (2024 model) and a Nissan e-note (2024 model). In this experiment, we targeted the Japanese 30 km/h speed limit sign: this sign is detected correctly by both our YOLOv5 model trained on the GTSRB dataset and the production TSR systems as shown in Fig.~\ref{fig:commercial-TSR}.
    Since the production TSR systems only display recognition results after passing each sign, we conducted 20 trials approaching the sign from 50 meters at 5 km/h for each vehicle and calculated ASR as the percentage of trials where the system failed to detect the sign.

    \noindent\textbf{Results.}
    Our baseline testing without the attack demonstrated reliable performance across both systems, with recognition rates of 95\% for the Toyota Yaris and 90\% for the Nissan e-note over 20 test runs each. However, when subjected to our attack, both systems' recognition rates dropped substantially: the Toyota Yaris dropped to 40\% (ASR of 60\%) and the Nissan e-note dropped to 25\% (ASR of 75\%). %
    Although these ASRs on commercial TSR systems (60-75\%) are lower than those against research-grade TSR models ($>$93\%), our attack maintains effectiveness against production systems while prior physical attacks on speed-limit signs were unable to succeed on commercial systems (0\% ASR)~\cite{wang2024revisiting}. 

    The reduced effectiveness compared to research-grade models likely reflects the temporal-memory mechanisms built into production systems, as suggested by recent studies~\cite{wang2024revisiting}. These systems temporally memorize a detection result for a very short time, making them more robust against single-frame attacks. In contrast, our superior performance over existing approaches can be attributed to the unique properties of retroreflective patches—they reflect light directly back toward the source regardless of the incident angle, ensuring consistent attack activation when illuminated by vehicle headlights. This persistent activation mechanism enables our attack to remain effective even against systems with temporal robustness.
    Overall, our results demonstrate that our ARP attack poses a realistic threat to production AD systems.

%% file: src/07-Defence.tex
\nsection{Defense}\label{sec:defense}

We first discuss existing defenses against patch attacks and their limitations, then introduce DPR Shield, our defense strategy leveraging retroreflection characteristics.

\nsubsection{Prior Defense Strategies for Patch Attacks}
    While ARP attacks appear natural to human observers, they function as adversarial inputs that intentionally cause misrecognition in TSR models.
    Therefore, existing defenses against adversarial patches can be theoretically applicable against our attack as well.
    So far, defenses against patch attacks are generally categorized into two types~\cite{sato2024invisible}: empirical defenses~\cite{DetectionGuard,ji2021adversarialyolodefensehuman,liu2022segmentcompletedefendingobject,hayes2018visible,naseer2019local,wu2019defending} and certified defenses with theoretical guarantees~\cite{xiang2022patchcleanser, xiang2022objectseekercertifiablyrobustobject}. Empirical defenses are generally known to exhibit lower robustness and be more vulnerable to adaptive attacks compared to certified defenses~\cite{chiang2020certifieddefensesadversarialpatches}. Therefore, we focus on the certifiable defenses, particularly PatchCleanser~\cite{xiang2022patchcleanser}, as it is the state-of-the-art for defending classifiers against adversarial patch attacks. Thus, we  evaluate its defense performance of PatchCleanser against ARP attack in Appendix~\ref{sec:patchcleanser-eval}.

    However, PatchCleanser fails against our ARP attack while substantially reducing accuracy on benign traffic signs (Appendix~\ref{sec:patchcleanser-eval}).
    This is because PatchCleanser defense performance relies on random masking strategies --- a common approach in certified defenses. While masking portions of an image works effectively for general classification tasks (such as distinguishing between cats and dogs), it could not hold for TSR tasks, where the entire region of the sign provides critical semantic information and masking some parts of them can cause loss of the information for classification.
    We thus are motivated to design a new defense against the ARP attack by leveraging the domain-specific knowledge.

\nsubsection{Proposed Defense Strategies: DPR Shield}

ARP attacks exploit retroreflective materials, so mitigating these reflections is a natural defense approach. Polarizing filters are effective against many reflection-based problems~\cite{envinsci2025,exploratoriumPolarizedSunglasses} because they selectively block polarized light—light waves that oscillate  in one direction. When natural unpolarized light reflects off certain surfaces like water, it becomes partially polarized in a predictable direction, allowing polarizing sunglasses to block the glare while preserving overall visibility.
However, single polarizing filters fail against retroreflective materials because retroreflection preserves the diverse polarization states of the original unpolarized incident light, making selective filtering impossible.

Our key insight is to exploit the fundamental difference in how retroreflective patches and normal sign surfaces handle polarized light. We propose DPR Shield: by intentionally creating polarized illumination at the headlight, we can selectively suppress adversarial reflections at the camera.

The mechanism operates as follows: The headlight polarizer creates controlled polarized light. When this polarized light hits adversarial retroreflective patches, the reflection preserves the original polarization state. The camera's perpendicularly-oriented filter then blocks this preserved polarization. In contrast, normal sign surfaces scatter the incident light through diffuse reflection, which depolarizes it—allowing the sign content to remain visible through the camera filter.
This controlled polarization approach enables differential filtering: adversarial retroreflections are suppressed while legitimate sign visibility is maintained, with no computational overhead.

The setup of the DPR Shield is as shown in Fig.~\ref{fig:defense-setup} in Appendix\ref{appendix:dpr_shield}.

        \begin{figure*}[t!]
        \center
        \includegraphics[width=\linewidth]{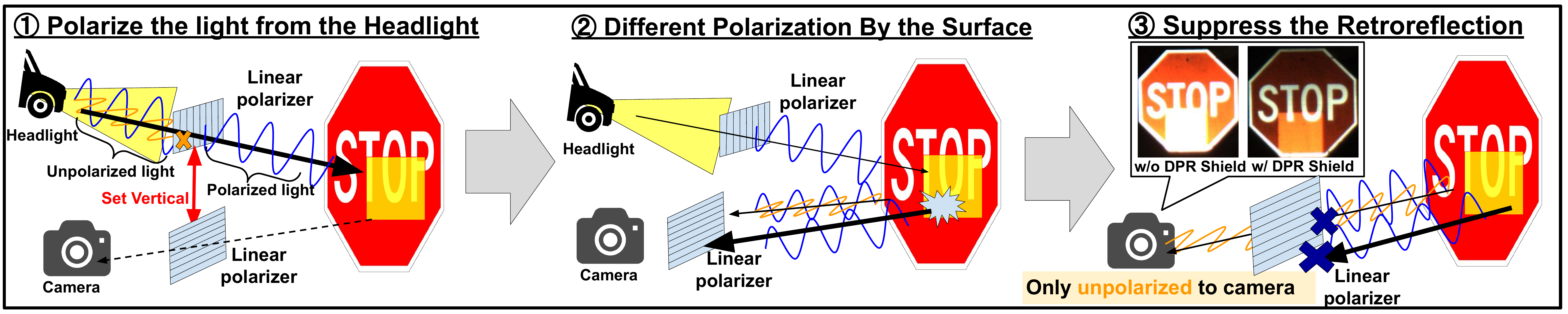}
  \vspace{-0.1in}
        \caption{Overview of polarization-based defense against retroreflective adversarial patches. The defense consists of three steps: (1) polarizing the headlight beam, (2) leveraging different polarization properties of sign and patch surfaces, and (3) suppressing retroreflective effects from adversarial patches while maintaining sign visibility.}
        \label{fig:DefenceMethod}
      
    \end{figure*}
\nsubsection{Evaluation}
\noindent\textbf{Experimental Setup.}
To evaluate the effectiveness of DPR Shield, we conducted real-world experiments as described in~\S\ref{sec:eval_real}, focusing on the scenario with the highest ASR: $d_{lon} = 15~\text{m}, d_{lat} = 0~\text{m}$. 
We used the same attack configurations (patch size, material) that demonstrated maximum effectiveness in our previous evaluation in \S\ref{sec:evaluation}. 

\noindent\textbf{Results.}
    DPR Shield maintained 100\% recognition accuracy on benign signs. Under attack conditions, effectiveness varied by sign type: For STOP signs, DPR Shield completely eliminated attacks (ASR: 100\% → 0\%) across both architectures. For SL65 signs, it provided complete protection for single-stage (ASR: 100\% → 0\%) but partial mitigation for two-stage architectures (ASR: 100\% → 25\%). Additionally, we observed patch darkening when using glass bead-based materials, we investigate this effect using STOP sign.

\noindent\textit{Evaluation of DPR Shield against Glass Bead-based Patches.}
    To further investigate the effectiveness of our defense mechanism against different types of retroreflective materials, we conducted additional experiments focusing on glass bead-based patches (NittoL). We evaluated the defense performance using STOP signs. 
    The patch configurations were optimized based on our previous findings --- for single-stage architecture and two-stage architecture, we optimize with $MPR$ of 0.25 that achieved the highest ASR.
    
    DPR Shield completely eliminated attacks (ASR: 100\% → 0\%) across both architectures. While the patches appeared darkened in images, this darkening did not compromise STOP sign visibility due to the low contrast against the red background. Although white backgrounds experience partially accuracy degradation due to higher contrast with darkened patches, red STOP signs remain unaffected. This demonstrates that DPR Shield can effectively defend even against glass bead-based patches without losing sign readability.

        \begin{figure}[t]
                \centering
            \includegraphics[width=\linewidth]{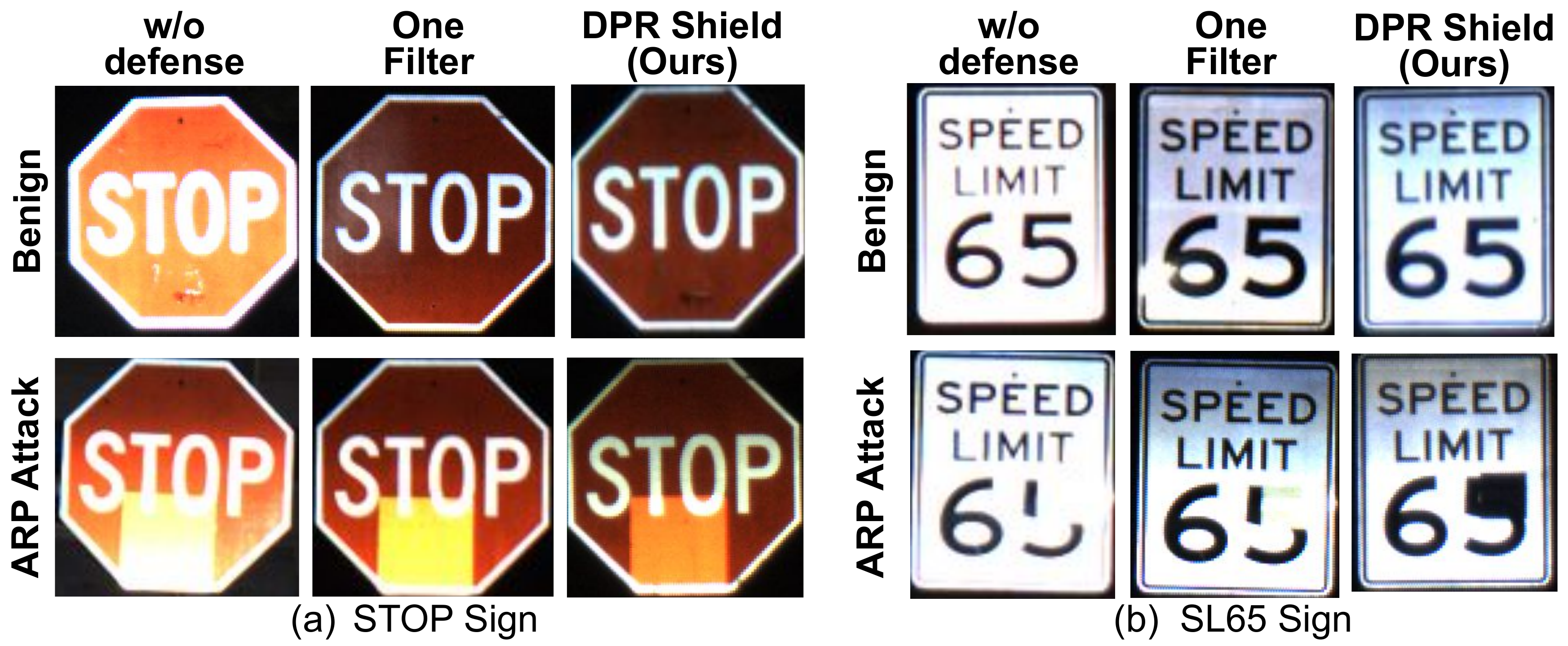}

                \caption{Comparison of defense strategies against ARP
attacks on STOP and SL65 signs.}
                \label{fig:defence-result-fig}
              
            \end{figure}

\begin{table}[t]
    \centering
    \footnotesize
    \setlength{\tabcolsep}{1.5pt}
 
        \setlength{\aboverulesep}{0pt}
        \setlength{\belowrulesep}{0pt}
    \renewcommand{\arraystretch}{1.0}
    \caption{
    TSR accuracy with and without our DPR Shield under benign and attack scenarios. The defense improves accuracy under attack without affecting benign performance.}

    \begin{tabular}{l cccc cccc}
        \toprule
        & \multicolumn{4}{c}{STOP} & \multicolumn{4}{c}{SL65} \\
        \cmidrule(lr){2-5} \cmidrule(lr){6-9}
        & \multicolumn{2}{c}{Benign} & \multicolumn{2}{c}{Attack} & \multicolumn{2}{c}{Benign} & \multicolumn{2}{c}{Attack} \\
        \cmidrule(lr){2-3} \cmidrule(lr){4-5} \cmidrule(lr){6-7} \cmidrule(lr){8-9}
        & Single & Two & Single & Two & Single & Two & Single & Two \\
        \midrule
        No Defense  & 100\% & 100\% & 0\% & 0\% & 100\% & 100\% & 0\% & 0\% \\
        With Defense & 100\% & 100\% & \textbf{100\%} & \textbf{100\%} & 100\% & 100\% & \textbf{100\%} & \textbf{75\%} \\
        \bottomrule
    \end{tabular}
    \label{tab:defense_comparison}

\end{table}

%% file: src/08-Discussion_Limitaiton.tex
\nsection{Discussions and Limitations}\label{sec:discuss}

\noindent\textbf{Limited Attack Effective Situation.}

While our study primarily focused on nighttime conditions, ARP attacks are effective in any environment with sufficiently low ambient light where headlights are typically used. STOP sign attacks demonstrated high efficacy ($\ge$85\% ASR) at ambient light levels below 900 lux, corresponding to conditions approximately 20 minutes before sunset, while SL65 sign attacks maintained effectiveness ($\ge$90\% ASR) even at illumination levels up to 4000 lux. These thresholds encompass numerous common driving scenarios beyond nighttime, including tunnels (approximately 54 lux), parking structures, overcast or rainy weather (approximately 100 lux), and dusk periods~\cite{SPIELuminousEfficacy}. Since drivers routinely activate headlights in these environments regardless of time of day, ARP attacks present a practical threat across a wide range of everyday driving conditions.

\noindent\textbf{Closed-Loop End-to-End Driving Evaluation.}
While comprehensive end-to-end evaluation using open-source AD stacks would provide additional insights, current platforms face practical limitations: Apollo and Autoware lack integrated TSR components, while OpenPilot's TSR functionality remains experimental and unstable. Our evaluation on commercial TSR systems (Toyota Yaris, Nissan eNote) provides direct evidence of real-world impact, as TSR misrecognition immediately affects longitudinal control for critical signs like STOP and speed limits—the primary safety-critical pathway in AD. This TSR-focused methodology aligns with established security research practices where component-level attacks demonstrate system-level vulnerabilities. Future work should explore attack propagation through complete AD pipelines as these systems mature and become more accessible for security research.

\noindent\textbf{Map-based Defense.}
While map-based defense might be theoretically feasible, it is not a practical solution due to several real-world constraints. Maintaining high-definition (HD) maps requires significant costs and limited coverage; even industry leader Waymo maintains maps only for extremely limited areas such as San Francisco~\cite{Waymo2025}. This reality is reflected in industry practice, where a recent survey showed that 12 out of 13 major automotive manufacturers rely exclusively on camera-based systems for their TSR functionality~\cite{wang2024revisiting}. Furthermore, map data has inherent limitations, including update delays and the inability to capture temporary signs. Given these constraints, vision-based recognition remains the primary and authoritative source for TSR. Therefore, ARP attack represents a realistic threat to production TSR systems.

\noindent\textbf{Adaptive Attacks.}
This research does not cover an adaptive attack against our DPR Shield because we first want to focus on investigating the attack capability of retroreflective patches on the existing TSR system. Our DPR Shield can effectively mitigate the ARP attack effect especially for stop signs because the light reflection is significantly suppressed. However, effective attack may exist even under low intense reflection since the adversarial attack can originally be effective with a very subtle perturbation. The attacker may combine different types of retroreflective materials to achieve it. We leave this arm racing for future research.

%% file: src/09-Conclusion.tex
\vspace{-0.05in}
\nsection{Conclusion} \label{sec:conclusion}

In this study, we discover a novel attack vector named ARP that leverages retroreflective patches to fool the TSR systems in AD systems. The ARP attack is activated only when the headlights of the victim AD vehicle shine them by the retroreflectivity that selectively reflects most incoming lights back to their source direction. This design allows the adversary to simultaneously achieve the advantages of the two major prior attacks: the deployability of patch attacks and the stealthiness of the laser or light projection attacks. 

To enable effective ARP attacks, we designed a methodology to simulate retroreflective light behavior using a 3D shading simulator and optimize attack parameters through black-box optimization to address the limitations of prior approaches.
Our evaluation demonstrates the attack's effectiveness, achieving 100\% ASR in static scenarios and $\geq$93.4\% ASR at ranges beyond 35m in dynamic driving scenarios with 5 km/h speed.
Moreover, we evaluate against a production TSR system in dynamic scenario and confirm that the attack achieve up to 75\% ASR.
User studies confirm that ARP patches appear as natural as benign signs (avg. score 2.04 vs 1.81).
To defend against this threat, we propose DPR Shield, a polarization-based defense that provides 100\% protection for single-stage TSR systems while maintaining 75\% effectiveness for two-stage architectures under attack conditions. 

\nsection{Acknowledgements}

This research was supported in part by NEDO JPNP25006, JSPS KAKENHI 22H00519,  JST CREST JPMJCR23M4, and JST BOOST, Japan Grant Number JPMJBS2429, USDOT under Grant 69A3552348327 for the CARMEN+ University Transportation Center, NSF under grants CNS-2145493 and CNS-2413877.

%% file: src/10-Appendix.tex
\appendix

\nsection{Detail of Threat Model}
    Fig.~\ref{fig:threat_model} illustrates an overview of the ARP attack threat model,
    which generally follows the same threat model as patch at-
    tacks. The major difference from the prior patch attacks is
    the prerequisite of the victim vehicle’s headlights to trigger
    the attack.

\begin{figure}[h]
        \centering
        \includegraphics[width=1\linewidth]{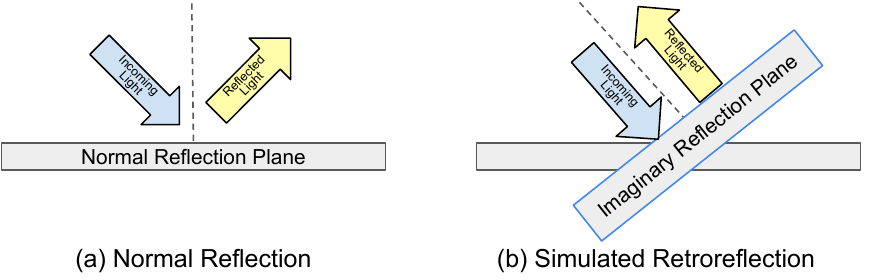}
        \caption{Simulated retroreflection with imaginary perpendicular reflection plane. As the retroreflection reflects lights back to their source, we can simulate it with specular reflection on the perpendicular plane against the light.}
        \label{fig:imaginary_plane}
    \end{figure}

\nsection{Detail of Imaginary Perpendicular Reflection Plane}\label{sec:appendix_imaginary_plane}
    The basic property of the retroreflective light reflection is to reflect the incoming light back to its source direction. However, we do not have such a reflection option in recent 3D shading simulators, which only have options for diffuse and mirror reflections. To simulate the retroreflectivity, we place an imaginary reflection plane perpendicular to the incoming light as described in Fig.~\ref{fig:imaginary_plane}.

\nsection{Four retroreflective patches on the stop sign}
\label{sec:appendix-patch-day-night}

\begin{figure}[t!]
    \centering
    \includegraphics[width=1\linewidth]{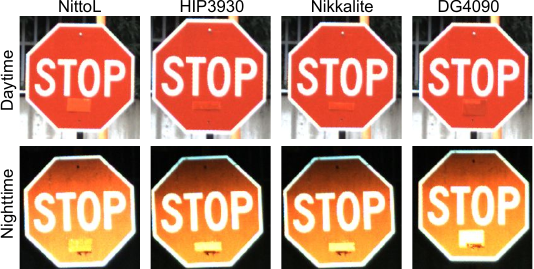}
    \vspace{-0.1in}
    \caption{Example of the four retroreflective patches on the stop sign. The reflection intensities of each patch are along with their reported retroreflective coefficients as in Table~\ref{table:selected_patch}.}
    \label{fig:four_patch_day_night}
    \vspace{-0.2in}
\end{figure}

We show the example images of the retroreflective patches on the stop sign in Fig~\ref{fig:four_patch_day_night}

\begin{figure}[h]
            \begin{minipage}{.49\linewidth}
            \centering
            \includegraphics[width=\linewidth]{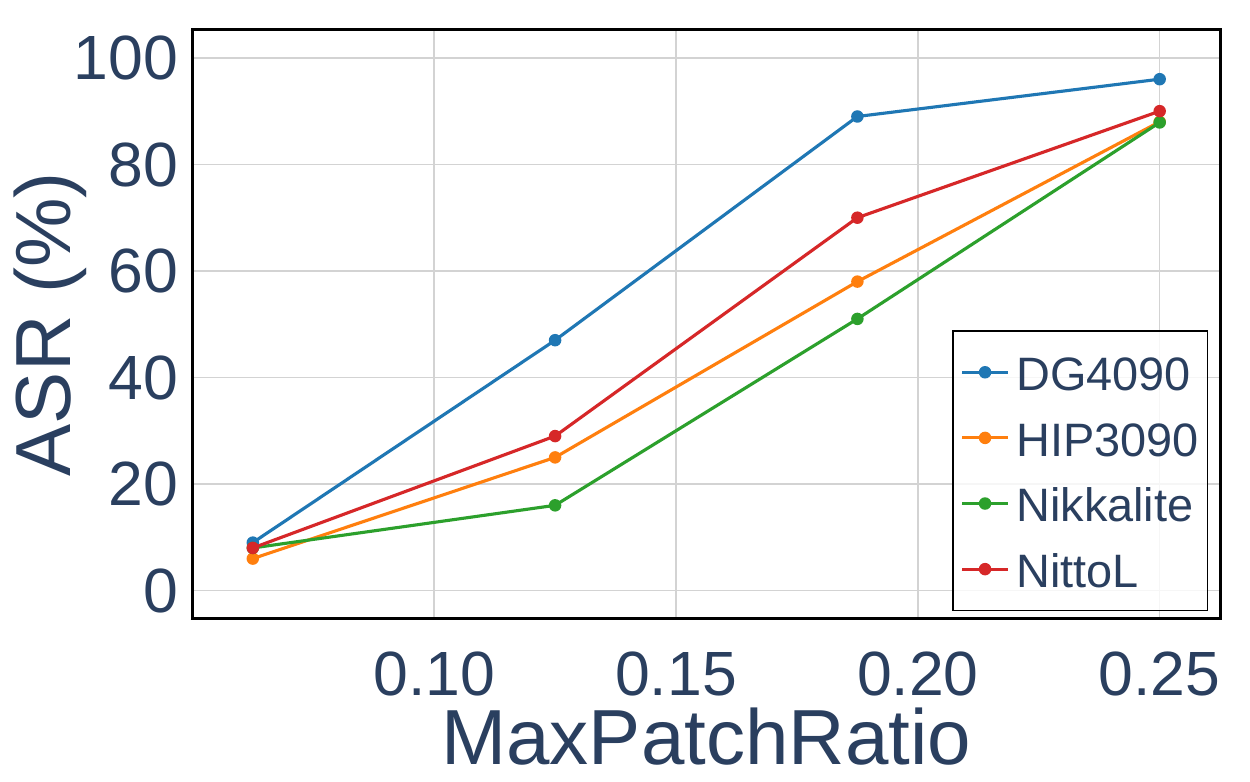} 
            \subcaption{STOP sign}
                   
          \end{minipage}
            \begin{minipage}{.49\linewidth}
                \centering
                \includegraphics[width=\linewidth]{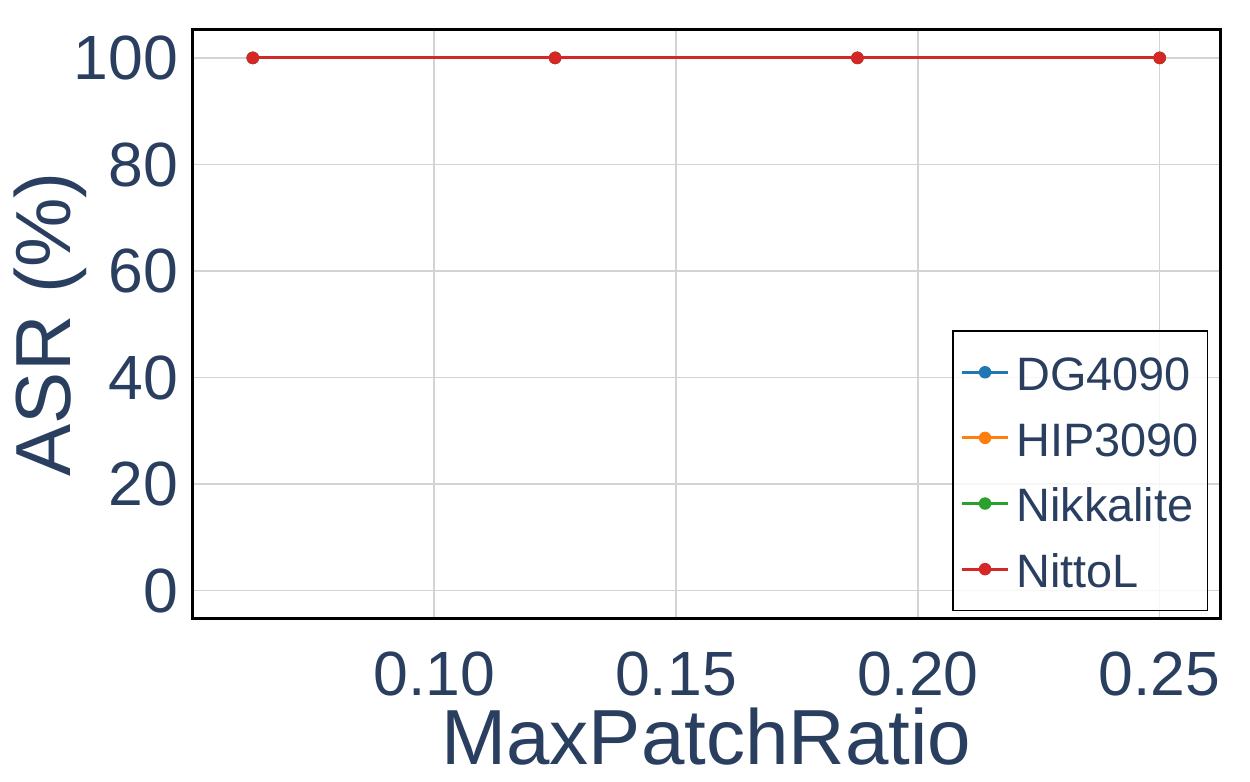}
                \subcaption{SL65 sign}

          \end{minipage}
          
          \caption{Impact of patch retroreflective materials and sizes on ASR against single-stage TSR for (a) STOP sign and (b) SL65 sign)
          }
          \label{Attack_YOLO}
        \end{figure}

\begin{figure}[h]
            \begin{minipage}{.49\linewidth}
            \centering
            \includegraphics[width=\linewidth]{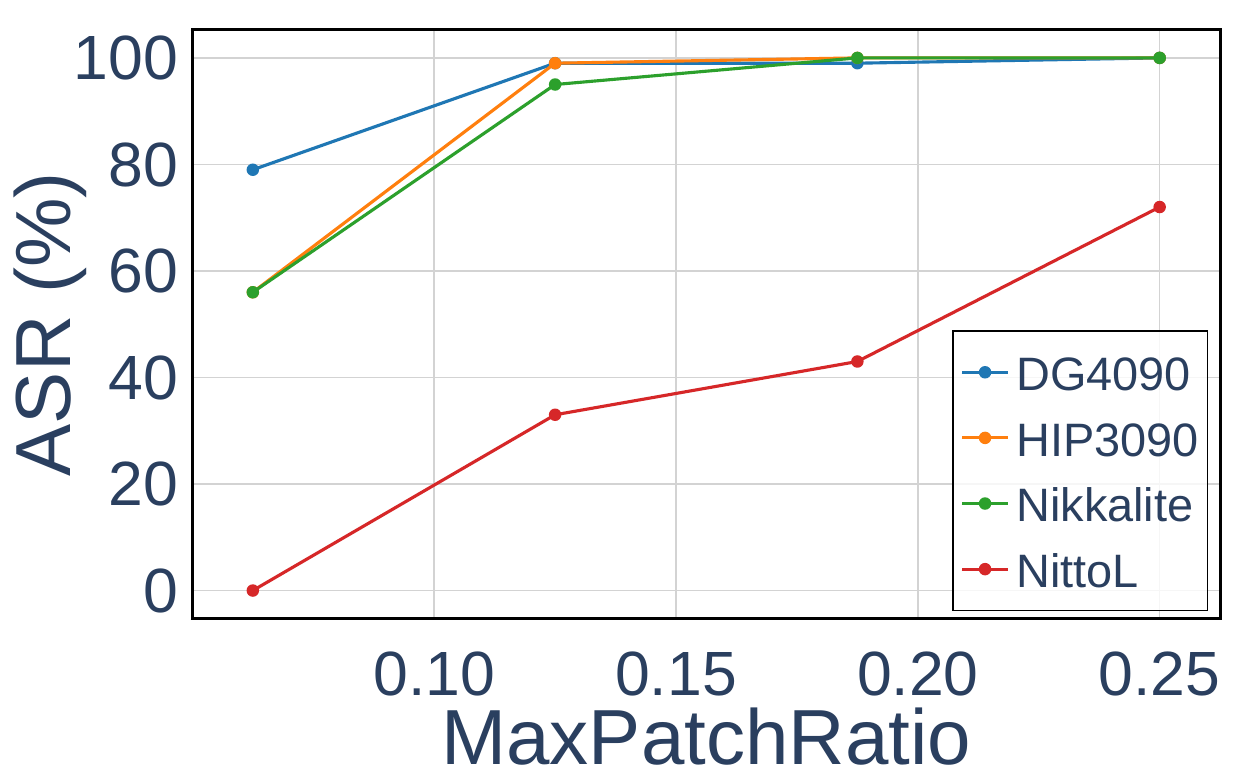}  
            \subcaption{STOP sign}        

          \end{minipage}
            \begin{minipage}{.49\linewidth}
                \centering
                \includegraphics[width=\linewidth]{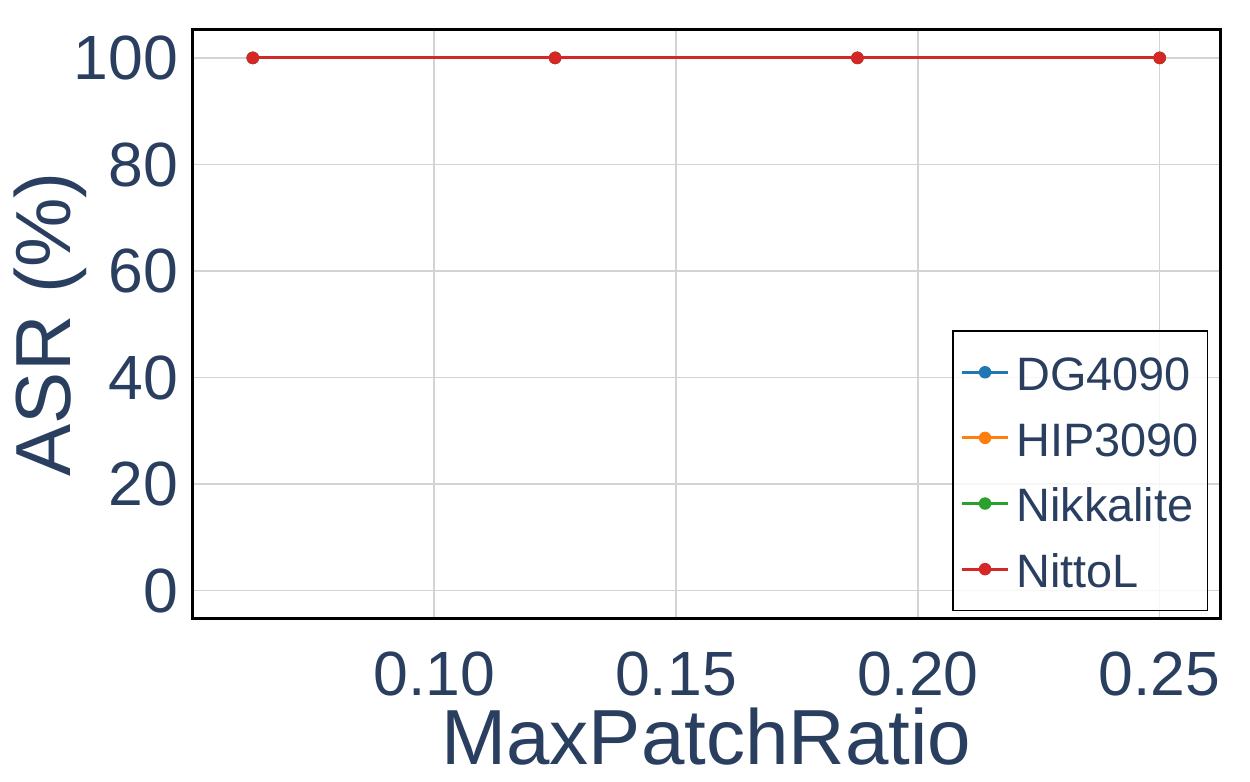}
                \subcaption{SL65 sign}

          \end{minipage}
        
          \caption{Impact of patch retroreflective materials and sizes on ASR against classifier in two-stage TSR for (a) STOP sign and (b) SL65 sign 
          }
           \label{Attack_CNN}
            \end{figure}

\nsection{Benign Performance of TSR Models in Our Research}\label{sec:appendix_benign-model-performance}
Table~\ref{table:model-performance} lists the benign performances of the
targeted single-stage and two-stage architectures models.
        \begin{table}[h]
        \vspace{0.2in}
        \centering
        \footnotesize
        \setlength{\aboverulesep}{0pt}
        \setlength{\belowrulesep}{0pt}
        \renewcommand{\arraystretch}{1.4}
        \caption{Benign Performance of the object detectors and classifiers for traffic sign recognition. 
        YOLOv5 is evaluated in mAP50. Others are in mAP.}
        \scalebox{0.9}{
              \begin{tabular}{ccccc}\toprule
                Object Detector (Training Dataset)     & mAP/APb &  &    Classifier (Training Dataset)         & Acc. \\ \cline{1-2} \cline{4-5} 
                YOLOv5~\cite{YOLO} (ARTS~\cite{ARTS_dataset})      & 57.0     &  & SimpleCNN (ARTS~\cite{ARTS_dataset})  & 81\% \\
                YOLOv5~\cite{YOLO} (COCO~\cite{lin2015microsoft}) & 37.2     &  & SimpleCNN (LISA~\cite{LISA_dataset})  & 99\% \\
                Faster R-CNN ~\cite{Faster-RCNN} (Mapillary~\cite{mapillary_dataset})           & 21.9     &  & SimpleCNN (GTSRB~\cite{GTSRB}) & 98\% \\
                Faster R-CNN ~\cite{Faster-RCNN} (ARTS~\cite{ARTS_dataset})           & 77.0     &  &  &  \\ \cline{1-2} \cline{4-5}  
              \bottomrule
              \end{tabular}
        }
        \label{table:model-performance}
        \vspace{-0.1in}
        \end{table}
\nsection{Detail of Attack Effectiveness Evaluation}\label{Append:Attackeffect}
    \noindent\textbf{Impact of Patch Size and Path Material on ASR.}
    We analyze the impact of path size and patch retroreflective material on the YOLOv5 trained on the ARTS dataset for single-stage architecture and SimpleCNN trained on ARTS dataset for two-stage architecture.
    We generate 100 attacks for each combination of four $MPR$ values (0.0625, 0.125, 0.1875, 0.25) and four retroreflective materials listed in Table~\ref{table:selected_patch}.
        Fig.~\ref{Attack_YOLO} and~\ref{Attack_CNN} show the ASR for each combination of the retroreflective materials and $MPR$.   
        The effects of $MPR$ and retroreflective materials on ASR showed different trends depending on the target TSR model architecture and sign type. For SL65 signs, the ARP attack achieves 100\% ASR against both single- and two-stage architectures for all the tested $MPR$. However, for STOP signs, the attack effectiveness varies between architectures. 
        We find that the DG4090 material, which has the highest retroreflectivity among tested materials, demonstrates the strongest attack performance for all $MPR$. The effects of patch size and materials on ASR showed distinctly different trends between STOP and SL65 signs. For SL65 signs, the ARP attack achieves 100\% ASR against both single- and two-stage architectures regardless of the $MPR$. This high effectiveness with small patches is likely due to two factors: the white background allows even small retroreflective perturbations to create significant contrast, and speed limit signs share similar visual features (rectanglar shape, numerical content) which makes misclassification between different speed limits easier.
        
        For STOP signs, however, the relationship between patch size and ASR varies between architectures. With DG4090 patches against single-stage architecture, the ASR gradually increases until reaching 100\% at $MPR$ of 0.1875, with no additional improvement beyond this size. Against two-stage architecture, a smaller $MPR$ of 0.125 is sufficient to achieve maximum effectiveness. The STOP sign requires larger patches likely because of its unique octagonal shape and distinctive text, which creates strong discriminative features that are harder to manipulate compared to SL65 signs.

    \noindent\textbf{Attack Impact of Multiple Patches.}
    We investigated the impact of dividing the attack area into multiple patches against two-stage architecture.
    For each $MPR$ value (0.0625, 0.125, 0.1875, 0.25), we divided the total allowed area into N equal-sized patches (N = 1, 2, 3, 4, 5), where each patch has a $MPR$ of the total allowed area divided by N. For example, with a total $MPR$ of 0.125 and N = 2, each patch would have a $MPR$ of 0.0625. The position of each patch was optimized using our methodology described in~\S\ref{sec:attack_generation}.
    As shown in Fig.~\ref{tab:patch_numbers}, the ASR consistently decreases as the number of patches increases, even when maintaining the same total attack area. This trend holds across all tested $MPR$ values. 

    These results indicate that using a single, larger patch is more effective than multiple smaller patches. 

        \begin{table}
        \setlength{\aboverulesep}{0pt}
        \setlength{\belowrulesep}{0pt}
        \setlength{\tabcolsep}{4pt}  
        \renewcommand{\arraystretch}{1.2}
        \footnotesize
        \centering
        \caption{ASR for varying numbers of patches under different $MPR$ constraints. $MPR$ is expressed as a ratio of the traffic sign's bounding box area.
        Bold numbers indicate the highest ASR for each total patch area ratio.}
        \label{tab:patch_numbers}
        \begin{tabular}{c|ccccc} 
        \hline
       $MPR$ & \multicolumn{5}{c}{Number of Patches} \\
        (Total) & 1 & 2 & 3 & 4 & 5 \\
        \hline
        0.0625 & \textbf{70.0\%} & 5.0\% & 0.0\% & 0.0\% & 0.0\% \\
        0.1250 & \textbf{95.0\%} & 90.0\% & 65.0\% & 40.0\% & 15.0 \%\\
        0.1875 & \textbf{100.0\%} & 95.0\% & 95.0\% & 75.0\% & 90.0\% \\
        0.2500 & \textbf{100.0\%} & \textbf{100.0\%}& \textbf{100.0\%} & \textbf{100.0\%} &\textbf{ 100.0} \\
        \hline
        \end{tabular}
    \end{table}

\nsection{Detail of Attack Transferability Evaluation}    \label{appendix_attack_trans}
    \noindent\textbf{Architecture Transferability.}
         \textit{Single-Stage Architectures.} 
        We evaluate the transferability between two representative object detection architectures: Faster R-CNN and YOLOv5 both trained on the ARTS dataset.  
        Our analysis revealed two distinct patterns in attack transferability. 
        First, the attack for a SL65 sign demonstrated consistent high transferability (100\% ASR) across both architectures.
        For STOP sign attacks, we observed dropped effectiveness from 90\% ASR to 55\% ASR.
        This asymmetric transferability pattern can be attributed to the architectural differences in feature extraction mechanisms. 
       Therefore, while our ARP attack demonstrates transferability from one model to another, this transferability may drop when the extracted features used for detection differ significantly between models. 

        \textit{Two-Stage Architectures.}
            We evaluate ARP attack transferability across three datasets (ARTS, GTSRB and LISA) using the same model (Simple CNN).
            Table~\ref{table:trans_dataset_classifier} shows that ARP attack maintains high transferability (65-100\% ASR) across most different training datasets for both STOP and Speed Limit signs.
            The reduced transferability of GTSRB-generated attacks to ARTS models can be attributed to the potential absence of successfully targeted misclassification classes in the ARTS dataset.
        
    \noindent\textbf{Dataset Transferability.}
        \textit{Single-Stage Architectures.}
            Table~\ref{tab:trans_dataset_detector} lists the ASR for the ARP attack transferred between different datasets (ARTS, Mapillary) using Faster R-CNN~\cite{Faster-RCNN}.  
            Our transferability analysis revealed different trend between STOP and SL65 sign attacks.
            For SL65 signs, the ARP attacks demonstrate robust transferability (40-100\% ASR) across the datasets. However, for STOP signs, while attacks achieve high effectiveness ($\geq$65\% ASR) within ARTS-trained models, they show no transferability to Mapillary-trained models.
            This result likely stems from the fundamental difference in dataset taxonomy: while ARTS contains multiple traffic sign classes, Mapillary is specifically limited to STOP and SL65 sign detection. This restricted class space in Mapillary appears to induce more robust feature representations for STOP sign detection, making the model less susceptible to the ARP attack optimized for detector trained on ARTS dataset.

        \textit{Two-Stage Architectures.}
           Table~\ref{table:trans_arch_classifiers_STOP} and Table~\ref{table:trans_arch_classifiers_SL65} list the ASR of transferred attacks for different model architectures. 
           For STOP signs, attacks generated using the Simple CNN architecture achieved high transfer ASR across other architectures (83-100\%), while those generated by ResNet showed limited transferability (25-40\%). 
           For SL65 signs, attacks demonstrated robust transferability across most architectural combinations, with success rates consistently exceeding 90\%.
           These results show that the ARP attack can be transferred from one model to another if the two models employ similar approaches to feature extraction, but may fail when their underlying feature representations differ substantially, as seen in the case of ResNet compared to other architectures.

        \begin{table*}[t!]
            \centering
            \footnotesize
            \setlength{\tabcolsep}{3pt}
            \setlength{\aboverulesep}{0pt}
            \setlength{\belowrulesep}{0pt}
            \renewcommand{\arraystretch}{1.1}
            \caption{Transferability evaluation results with ASR for different object detectors. Bold numbers indicate ASR above 80\%.}
            \label{table:trans_arch_detectors}
            \begin{tabular}{cccccccc}
            \toprule
             &               & \multicolumn{6}{c}{Target model}                                                                               \\ \cline{3-8} 
             &               & \multicolumn{2}{c}{STOP Sign}           &  &                         & \multicolumn{2}{c}{SL65  Sign}                    \\ \cline{3-4} \cline{7-8} 
             &               & Faster R-CNN (ARTS) & YOLOv5 (ARTS) &  &                         & Faster R-CNN (ARTS) & YOLOv5 (ARTS) \\ \cline{1-4} \cline{6-8} 
            \multirow{2}{*}{\begin{tabular}[c]{@{}c@{}}Source\\ Model\end{tabular}} & Faster R-CNN (ARTS) & \textbf{90\%}& {55\%}& \multirow{2}{*}{} & Faster R-CNN (ARTS) & \textbf{100\% }& \textbf{100\% }\\
             & YOLOv5 (ARTS) & {55\%}& \textbf{90\%}&  & YOLOv5 (ARTS) & \textbf{100\% }& \textbf{100\% }\\ \bottomrule
            \end{tabular}
        \end{table*}
        
        \begin{table}[t]
            \centering
            \footnotesize
            \setlength{\tabcolsep}{3.0pt}
            \setlength{\aboverulesep}{0pt}
            \setlength{\belowrulesep}{0pt}
            \renewcommand{\arraystretch}{1.1}
            \caption{Transferability to different architectures of the classifier in the two-stage architecture for STOP sign.}
            \label{table:trans_arch_classifiers_STOP}
            \begin{tabular}{cc|cccc}
            \toprule
                                                                                    &               & \multicolumn{4}{c}{Target Model}                  \\
                                                                                    &               & SimpleCNN & Resnet50 & Dense Net & Efficient Net \\ \hline
            \multirow{4}{*}{\begin{tabular}[c]{@{}l@{}}Source\\ Model\end{tabular}} & SimpleCNN    &     99\%      &    100\%      &     87\%     &       83\%         \\
                                                                                    & Resnet50      &        25\%    &  100\%      &   40\%        &      30\%         \\
                                                                                    & Dense Net     &       40\%     &    65\%      &     100\%      &      70\%         \\
                                                                                    & Efficient Net &      55\%      &    70\%      &      65\%     &          100\%     \\ \toprule
            \end{tabular}
        \end{table}
        
        \begin{table}[t]
            \centering
            \footnotesize
            \setlength{\tabcolsep}{3.0pt}
            \setlength{\aboverulesep}{0pt}
            \setlength{\belowrulesep}{0pt}
            \renewcommand{\arraystretch}{1.1}
            \caption{Transferability to different architectures of the classifier in the two-stage architecture for SL65 sign.}
            \label{table:trans_arch_classifiers_SL65}
            \begin{tabular}{cc|cccc}
            \toprule
                                                                                    &               & \multicolumn{4}{c}{Target Model}                  \\
                                                                                    &               &  SimpleCNN & Resnet50 & Dense Net & Efficient Net \\ \hline
            \multirow{4}{*}{\begin{tabular}[c]{@{}l@{}}Source\\ Model\end{tabular}} & SimpleCNN    & 93\%       & 42\%     & 92\%      & 81\%          \\
                                                                                    & Resnet50      & 90\%      & 65\%     & 95\%     & 95\%         \\
                                                                                    & Dense Net     & 95\%      & 40\%     & 95\%     & 90\%         \\
                                                                                    & Efficient Net & 55\%       & 40\%     & 60\%      & 60\%          \\ \toprule
            \end{tabular}
            \vspace{-0.1in}
        \end{table}

        \begin{table}[t]
        \centering
        \footnotesize
        \setlength{\tabcolsep}{4pt}
        \setlength{\aboverulesep}{0pt}
        \setlength{\belowrulesep}{0pt}
        \renewcommand{\arraystretch}{1.1}
        \caption{Dataset transferability for STOP and SL65 sign with ASR against the single-stage TSR.}
        \label{tab:trans_dataset_detector}
        \begin{tabular}{cccc}
        \toprule
        & & \multicolumn{2}{c}{Target Model} \\ \cline{3-4}
        Sign Type & Source Model & \begin{tabular}[c]{@{}c@{}}Faster R-CNN\\(ARTS)\end{tabular} & \begin{tabular}[c]{@{}c@{}}Faster R-CNN\\(Mapillary)\end{tabular} \\ \midrule
        \multirow{2}{*}{STOP} & Faster R-CNN (ARTS) & \textbf{90\%}& 0\% \\
        & Faster R-CNN (Mapillary) & 65\%& 0\%\\ \midrule
        \multirow{2}{*}{SL65} & Faster R-CNN (ARTS) & \textbf{100\% }& 55\%\\
        & Faster R-CNN (Mapillary) & \textbf{100\% }& 40\%\\ \bottomrule
        \end{tabular}
        \end{table}

        \begin{table}[t]
        \centering
        \footnotesize
        \setlength{\tabcolsep}{5.5pt}
        \setlength{\aboverulesep}{0pt}
        \setlength{\belowrulesep}{0pt}
        \renewcommand{\arraystretch}{1.0}
        \caption{Dataset transferability with ASR against two-stage TSR.}
        \begin{tabular}{ccccc}
        \toprule
         & & \multicolumn{2}{c}{Target Dataset} & \\ \cline{3-4}
        Sign Type & Source Dataset & ARTS & GTSRB/LISA \\ \midrule
        \multirow{2}{*}{STOP} & ARTS & \textbf{99\%} & \textbf{94\%} \\
         & GTSRB & 0\% & \textbf{100\% } \\ \midrule
        \multirow{2}{*}{SL65} & ARTS & \textbf{100\% } & \textbf{99\% } \\
         & LISA & 65\% & \textbf{100\%} \\ \bottomrule
        \end{tabular}
        \label{table:trans_dataset_classifier}
        \end{table}

\nsection{Objective Stealthiness Evaluation with the Perceptual Similarity}\label{sec:setalth_perceptual}
    
    We objectively and quantitatively evaluate the stealthiness of the ARP attacks with widely recognized perceptual similarity metrics SSIM and LPIPS, which provide computational measurements as adopted for stealthiness evaluation in prior adversarial attack studies~\cite{EnhanceStealthiness_Xia_2024,li2023adv}.
    These metrics provide objective measurements of the visual discrepancies between the original and adversarial images. The SSIM only focuses on the image luminance and contrast but does not take image semantics into account well. The LPIPS is calculated by the similarity in a feature space embedded by a neural network and is expected to take semantics information into the similarity~\cite{zhu2024evaluate}. We make both SSIM and LPIPS range from 0 (no similarity) to 1 (perfect similarity.). In our context,  higher perceptual similarity scores means higher stealthiness.
    
    \noindent\textbf{Experimental Setup.}
    {
    As shown in Table~\ref{table:stealthiness-evaluation}, we calculate the perceptual similarities between the benign and 4 attack images on the stop sign. For the ARP attack, we generate attacks for both the single-stage and two-stage TSR systems. For the single-stage TSR, we use YOLOv5 trained on ARTS dataset ~\cite{ARTS_dataset}.
    For the two-stage TSR, we focus on the second-stage classifier and manually crop the stop sign area. We compare the ARP attacks with prior patch attacks, named RP$^2${1}~\cite{DBLP:journals/corr/abs-1807-07769} and RP$^2$\circled{2}~\cite{adv-patch} as baselines for comparison because:(1) they target the same traffic sign type (STOP sign) as our work, and (2) they have demonstrated deployability in physical-world settings through both poster-printing and sticker-based implementations.. We reproduced these attacks on our stop sign with patch images extracted from their papers.
    we utilize ARP patches crafted using DG4090 retroreflective materials with specific configurations, as described in \S~\ref{sec:eval_real}.
    These patches are designed to balance attack effectiveness and stealthiness in both single-stage and two-stage TSR systems.}
    
    \noindent\textbf{Results.}
    {
    As listed in Table~\ref{table:stealthiness-evaluation}, the ARP attacks show much higher stealthiness than prior attacks in both SSIM and LPIPS. The improvements are $\geq$0.1 and 0.15 points for SSIM and LPIPS, respectively. Our results show that the ARP attack is more stealth than the patch attacks which designed in prior works.
    }

        \newlength{\imgwidth}
        \setlength{\imgwidth}{1.2cm}  %
        
        \begin{table}[t!]
        \centering
        \footnotesize
        \setlength{\tabcolsep}{2.0pt}
        \caption{Perceptual-Similarity Scores of the benign, ARP attacks, and prior patch attacks.}
        \label{table:stealthiness-evaluation}
        \begin{tabular}{c|c|cc|cc}
        \hline
                      & Benign & \multicolumn{2}{c|}{ARP Attack (Ours)} & \multicolumn{2}{c}{Prior Attack} \\
                      &        & \scalebox{0.9}{Single-Stage} & \scalebox{0.9}{Two-Stage}  &      RP$^2$\circled{1}          &    RP$^2$\circled{2}            \\ \toprule
        Image &\includegraphics[width=\imgwidth]{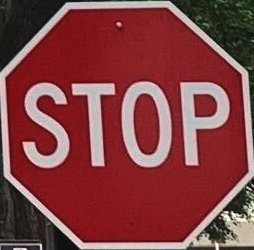} &
                       \includegraphics[width=\imgwidth]{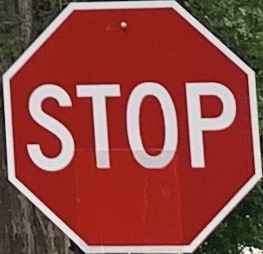} &
                       \includegraphics[width=\imgwidth]{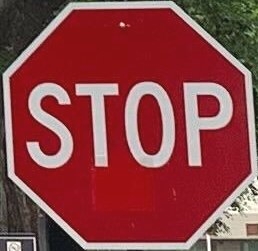} &
                       \includegraphics[width=\imgwidth]{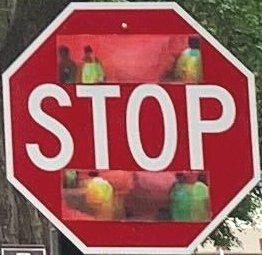} &
                       \includegraphics[width=\imgwidth]{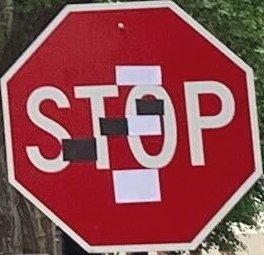} \\ \hline
        SSIM          & 1.0    & \textbf{0.60} & 0.59             & 0.47           & 0.50           \\
        LPIPS         & 1.0    & \textbf{0.91} & \textbf{0.91}            & 0.76           & 0.76           \\ \toprule
        \end{tabular}
        \end{table}

    \nsection{Subjective Stealthiness Evaluation with User Study}\label{Append:stealth-user}

    We conducted a user study to evaluate the subjective stealthiness of the ARP attack, asking human participants to assess the perception (natural or unnatural) of presented benign and attack images, as illustrated in Fig.\ref{fig:user_study_overview}. 
    The detailed experimental design and methodology are described in Appendix\ref{Append:stealth-user}.
        
    \noindent\textbf{Experimental Setup.}
            To evaluate the perceptual stealthiness of ARP attacks, we conducted a user study through Prolific~\cite{Prolific}, an online recruitment platform.  We recruited 50 participants (23 male, 23 female, 4 non-specified) with normal or corrected-to-normal vision. Participants assessed 60 traffic sign images across five conditions: unmodified signs, two variants of our ARP attack, and two existing attack methods (RP$^2$ ~\cite{DBLP:journals/corr/abs-1807-07769} and RP$^2$~\cite{adv-patch}). We created four viewing scenarios for each condition by combining temporal (day/night) and spatial factors (driver view at 2.0~m left, pedestrian view at 0.5~m right) at three different distances (10~m, 15~m, 20~m). Using a web-based interface optimized for desktop displays (minimum 1920×1080 resolution), participants rated each randomly-presented image's naturalness on a 5-point Likert scale, responding to the statement ``This traffic sign appears natural''. After applying an attention-check filter, we retained 37 valid responses for analysis, with each participant compensated £4.5 based on pilot study completion times and aligned with Prolific's recommended hourly rates.

        \begin{figure}[t]
        \includegraphics[width=\linewidth]{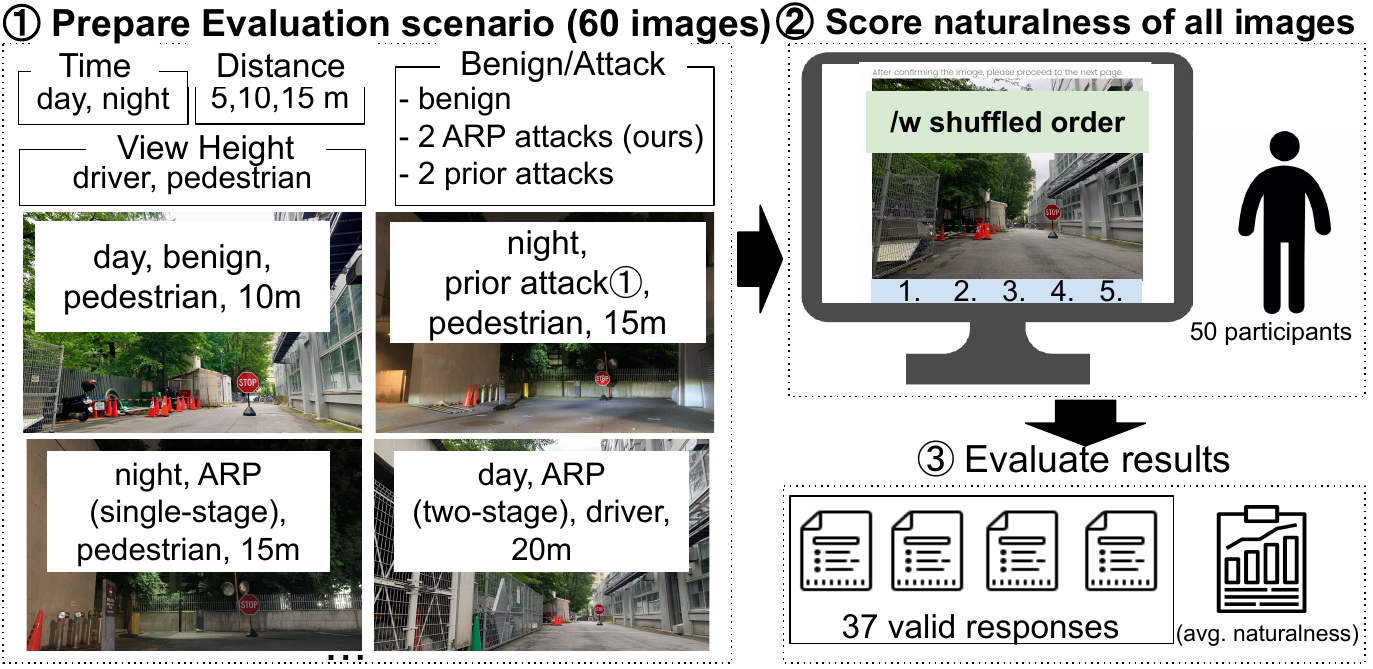}
        \caption{Overview of our stealthiness user study procedure. We (1) prepare 60 evaluation scenarios with different environmental factors and attack types, (2) then present all scenarios to 50 participants with shuffled orders for each, and (3) finally evaluate 37 valid responses after applying attention checks.}
        \label{fig:user_study_overview}
        \end{figure}

    \begin{table}[tbp]
        \centering
        \footnotesize
        \setlength{\tabcolsep}{2.5pt}
        \renewcommand{\arraystretch}{0.65}
        \caption{User Perception Scores (1--5). Lower scores indicate higher naturalness; higher scores indicate higher unnaturalness.Top 2 highest-rated attacks are in bold.}

        \label{tab:userstudy-results}
        \begin{tabular}{llc|ccccc}
        \toprule
        \multirow{2}{*}{Time} & \multirow{2}{*}{View} & \multirow{2}{*}{$d_{lon}$} & \multicolumn{5}{c}{Average Score} \\
        \cmidrule{4-8}
         &  &  & Benign & \multicolumn{2}{c}{Prior} &  \multicolumn{2}{c}{ARP  (Ours)} \\
          &  &  &  & RP$^2$\circled{1}  & RP$^2$\circled{2} & Single-Stage & Two-Stage \\
        \midrule
        \multirow{6}{*}{Day} & \multirow{3}{*}{Driver} 
         & 10m & 1.91 & 4.09 & 4.21 & \textbf{1.88} & \textbf{1.85} \\
         & & 15m & 1.82 & 4.15 & 4.38 & \textbf{1.94} & \textbf{1.85} \\
         & & 20m & 1.68 & 4.06 & 4.29 & \textbf{1.71} & \textbf{1.74} \\
        \cmidrule{2-8}
         & \multirow{3}{*}{Ped.} 
         & 10m & 2.09 & 3.82 & 4.47 & \textbf{2.06} & \textbf{2.03} \\
         & & 15m & 1.85 & 3.82 & 4.35 & \textbf{1.79} & \textbf{2.29} \\
         & & 20m & 1.74 & 3.82 & 2.94 & \textbf{1.85} & \textbf{1.85} \\
        \midrule
        \multirow{6}{*}{Night} & \multirow{3}{*}{Driver}
         & 10m & 1.82 & 3.68 & 4.47 & \textbf{1.91} & \textbf{1.94} \\
         & & 15m & 1.82 & 3.65 & 4.18 & \textbf{1.82} & \textbf{2.18} \\
         & & 20m & 1.76 &\textbf{1.94} & 3.97 & \textbf{2.21} & 2.82 \\
        \cmidrule{2-8}
         & \multirow{3}{*}{Ped.}
         & 10m & 1.71 & 3.74 & 4.41 & \textbf{2.18} & \textbf{2.24} \\
         & & 15m & 1.74 & 3.74 & 4.29 & \textbf{2.06} & \textbf{1.88} \\
         & & 20m & 1.76 & 3.82 & 4.26 & \textbf{1.91} & \textbf{1.79} \\
         \midrule
        \multicolumn{3}{c|}{Average}&1.81&3.69&4.19&\textbf{1.94}&\textbf{2.04}\\
        \bottomrule
        \end{tabular}

        \end{table}

\noindent\textbf{Results and Analysis.}
    Table~\ref{tab:userstudy-results} presents the average perception ratings for different scenarios and attack methods, based on participants' responses using the 5-point Likert scale, where lower/higher scores indicate higher perceived naturalness/unnaturalness. Our ARP attack consistently achieved high naturalness ratings (mostly between 1.71 and 2.29) across various scenarios, distances, and times of day, suggesting that participants frequently agreed or somewhat agreed that these modified signs appeared natural. This is particularly evident when compared to benign (unmodified) signs, which received similar ratings (between 1.68 and 2.09). In contrast, existing attack techniques (RP$^2$\circled{1} and RP$^2$\circled{2}) consistently received higher unnaturalness ratings (in many cases above 3.5).
    Furthermore, our ARP attack maintained high naturalness ratings even during nighttime scenarios when the attack was actively functioning by headlight reflection. 
    This can be attributed to our attack's utilization of retroreflection—a natural physical phenomenon—rather than specific artificial patterns used in conventional adversarial patches. Retroreflective materials are widely used in safety equipment, and protective clothing, making such reflections a common phenomenon in nighttime traffic environments. Consequently, retroreflective patches applied to signs are less likely to be perceived as unnatural, allowing ARP attack to maintain stealthiness even when attacks are triggered.

    The results also revealed that viewing perspective and distance influenced perception. The pedestrian perspective generally yielded lower naturalness ratings compared to the driver's perspective, particularly at closer distances. The distance $d_{lat}$ had a substantial impact - longer distances generally resulted in higher naturalness ratings. There was only one exception where the ARP attack received worse naturalness ratings than RP$^2$\circled{1}: from the driver's perspective at 20 m during nighttime. While RP$^2$\circled{1}'s patches became nearly invisible at 20 m, our ARP attack's high-intensity retroreflection remained visible, leading to lower naturalness scores.

    \nsection{Details of User Study Design}
    \label{Append:stealth-user-detail}
    \noindent\textbf{Study Design and Participant Recruitment.}
        This study involved 50 participants recruited through Prolific, an online platform known for its diverse and reliable participant pool. This initial participant group consisted of 23 males, 23 females, and 4 individuals who identified as other genders, providing a balanced and diverse sample for our analysis.
        All participants reported normal or corrected-to-normal vision. 
        Compensation for each participant was set at £4.5, a rate determined based on the average task completion time observed in our pilot study and aligned with the default payment standards recommended by Prolific~\cite{Prolific}. 
        This approach ensured fair compensation while maintaining the integrity of the study.
        To ensure consistent and high-quality visual presentation of stimuli, participants were explicitly instructed to use personal computers rather than smartphones or tablets. This requirement was implemented to ensure sufficient screen resolution and to maintain a standardized viewing experience across participants. The minimum required screen resolution was set at 1920×1080 pixels, which is commonly available on modern desktop and laptop displays.
    
    \noindent\textbf{Experimental Protocol and Image Analysis Procedure.}
        Our experimental design employed a within-subjects approach in which each participant was exposed to all conditions in a randomized order to mitigate potential order effects. We used a single, simultaneous, continuous procedure for rating the traffic sign images. The protocol for each trial was as follows: participants were presented with an image 
        Following the image presentation, participants were asked to rate the naturalness of the image using a 5-point Likert scale. The scale ranged from ``Strongly Agree'' (1) to ``Strongly Disagree'' (5) in response to the statement ``{\em This traffic sign appears natural.}'' 
        additionally, we used this 5-point Likert scale to measure the level of perception, with 1 indicating ``Very natural'' and 5 indicating ``Very unnatural.'' This score serves as a measure of the perception scale.
    
    \noindent\textbf{Stimulus Set and Viewing Conditions.}
        The stimulus set was carefully designed to encompass a wide range of scenarios and attack conditions. We prepared images to evaluate five different conditions: benign (unmodified) traffic signs, two variants of our proposed attack method, and two existing attack methods (A~\cite{adv-patch} and B~\cite{DBLP:journals/corr/abs-1807-07769}) for comparison. For each of these conditions, we simulated four different scenarios: nighttime and daytime perspectives from both a driver's and a pedestrian's perspective. The pedestrian's perspective was simulated at 0.5 m to the right of the sign, while the driver's perspective was simulated at 2.0 m to the left.
        To account for the effect of distance on perception, we prepared images at three different distances from the sign: 10m, 15m, and 20m. This comprehensive approach resulted in a total of 60 images (5 conditions × 4 scenarios × 3 distances). The order of image presentation was predetermined through a randomization process and consistently applied across all participants.

    \noindent\textbf{Data Quality Assurance.}
        To ensure the integrity of our data, we implemented quality control measures. A single attention check was built into the study design, in which participants were asked to identify unnatural elements in a benign (unmodified) traffic sign image. Failure to respond correctly to this attention check resulted in exclusion from the final analysis. In addition, we enforced a minimum time threshold for image viewing to ensure that participants gave each stimulus adequate attention.
    
    \noindent\textbf{Demographic Distribution.}
        Of the initial 50 participants, 13 were excluded from the final analysis, primarily due to failure of the attention check. This screening process ensured that only high-quality, attentive responses were included in our final data set.
        This carefully designed and executed user study provided us with robust data to evaluate the stealthiness of our proposed attack method compared to existing techniques in various realistic scenarios and viewing height.
        
\begin{figure}[h]
    \centering
    \includegraphics[width=0.5\textwidth]{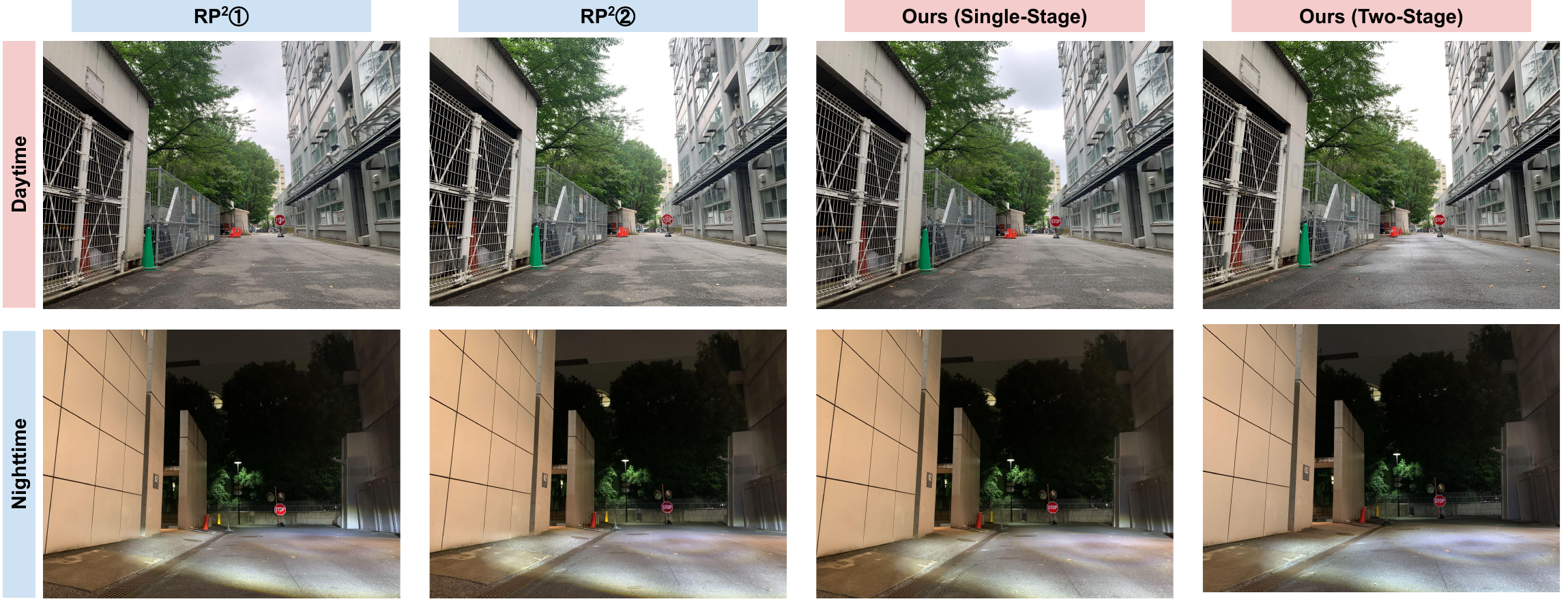}
    \caption{
Examples of images in user studies}
\label{fig:user-study-example-images}
    \end{figure}

\begin{figure}[h]
    \centering
    \includegraphics[width=0.5\textwidth]{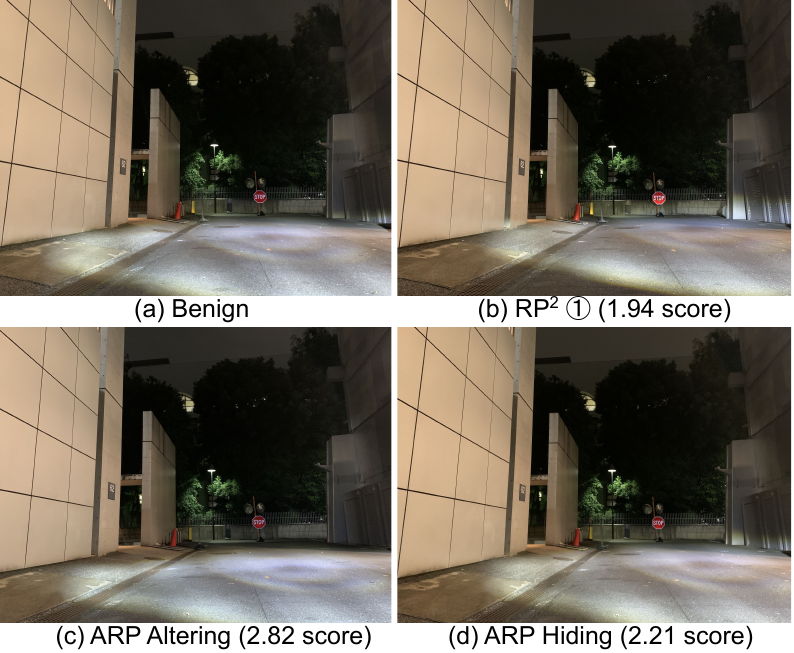}
    \caption{Example of the case that the RP$^2$\circled{1} has higher naturalness score than ARP attacks in the scenario from the driver's view at 20 m away in the nighttime.}
    \label{fig:worst-case}
    \end{figure}

    \noindent\textbf{Example Images Used in User Study.}
    Fig.~\ref{fig:user-study-example-images} illustrates example images used in our user study. Fig.~\ref{fig:worst-case} shows cases from \S~\ref{sec:eval_stealth} where prior attacks achieved higher naturalness scores compared to our ARP attack.
    
    \begin{figure}[h]
    
        \centering
        \includegraphics[width=0.5\textwidth]{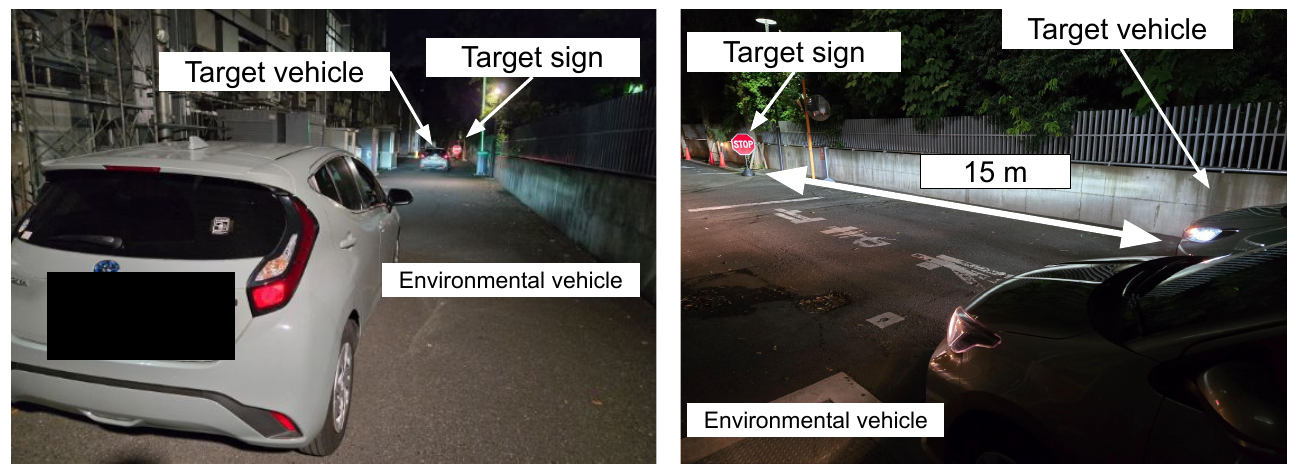}

          \caption{Experimental setup for multiple vehicle experiments. On the right is a situation where there is another vehicle behind the target vehicle, and on the left is a situation where there is another vehicle to the right of the target vehicle. Both target vehicle and environment vehicle turn on the high beam. }
           \label{fig:multiple-vehicles}
    \end{figure}
    
    \nsection{Detail of Physical Evaluation}\label{appendix_robust_sign_camera}
\noindent\textbf{Robustness to Different Sign Heights.}
    To evaluate our ARP attack robustness, we tested the ASR against single- and two-stage architectures across various sign heights ($h_s$ = 1.5, 1.75, 2.00 m). Attacks against the single-stage architecture achieved 100\% ASR across all heights. For two-stage architectures, STOP sign attacks achieved 76\% and 78\% ASR at heights of 1.75 m and 2.00 m respectively, while SL65 sign attacks maintained 100\% and 78\% ASR. This robustness stems from the retroreflective material's consistent ability to return light toward its source, even as viewing angles change with height.

\noindent\textbf{Robustness to Different Camera Positions.}
    To evaluate the spatial robustness of our attack, we examined its performance across various camera positions while maintaining the original headlight position. We tested two types of camera displacements: vertical shifts of 25~cm and 50~cm above the headlight position, and a horizontal shift of 25~cm to the left.
    The attack demonstrated consistent performance against Speed Limit sign detection, achieving a 100\% ASR across all test cases. This perfect success rate was maintained regardless of camera position and detection architecture, indicating that our attack method is particularly effective against Speed Limit sign recognition.
    However, the attack's effectiveness against STOP sign detection varied significantly depending on the underlying detection architecture. For the single-stage detector, the attack maintained robust performance across different camera positions. Vertical displacements of both 25 cm and 50 cm resulted in a 100\% ASR, while the horizontal displacement of 25 cm yielded a slightly reduced but still substantial ASR of 96\%.
    In contrast, the two-stage detection architecture demonstrated considerably higher robustness against our attack when the camera position was altered. With a 25 cm vertical displacement, the ASR dropped dramatically to 45\%. When the camera was raised by 50~cm or shifted horizontally by 25~cm, the attack completely failed, resulting in a 0\% ASR. These findings suggest that two-stage detection architectures possess inherent defensive properties against our headlight-based adversarial attacks when the camera position deviates from the attack's assumed position, particularly in the context of STOP sign detection.

\nsection{Attack Robustness Evaluation with Multiple Vehicles}\label{sec:appendix-multiple-vehicle}
    To evaluate the attack performance of the ARP attack under realistic traffic conditions, we replicated common intersection scenarios where multiple vehicles approach a STOP sign simultaneously in the real-world. 
        Our experimental setup replicated typical intersection conditions, with the target vehicle positioned 15 m from the STOP sign - a distance we chose to match previous physical patch attack studies for direct comparison. We then evaluated two common traffic scenarios: in the first, we positioned a second vehicle 7~m behind the target (22 m from the sign), and in the second, we placed a vehicle adjacent to the target with a 2~m lateral offset, both scenarios illustrated in Fig.~\ref{fig:multiple-vehicles}. Throughout our testing, all vehicles maintained active headlights to simulate realistic nighttime driving conditions.
        The ARP attack demonstrate that our ARP attacks are effective in multiple-vehicle scenarios, achieving 100\% ASR against target vehicles.
        These results suggest that ARP attacks remain effective in real traffic conditions where other vehicles are present.

    We positioned the target vehicle 15 m from the sign, matching the evaluation distance used in previous physical patch attacks to enable direct comparison. Additional vehicles were positioned at 15 m and 22 m behind and laterally offset from the target vehicle, simulating typical traffic queuing patterns at intersections.
    Results demonstrate that the ARP attack maintained 100\% ASR under all tested multi-vehicle configurations. This robust performance in realistic traffic scenarios suggests two key findings: First, our attack remains effective even when multiple vehicles are present, confirming its viability in real-world deployment conditions. Second, the directional nature of retroreflective materials ensures that the attack only affects the intended target vehicle while avoiding unintended triggering by other vehicles' headlights.

\begin{figure}[t!]
    \centering
    \includegraphics[width=\linewidth]{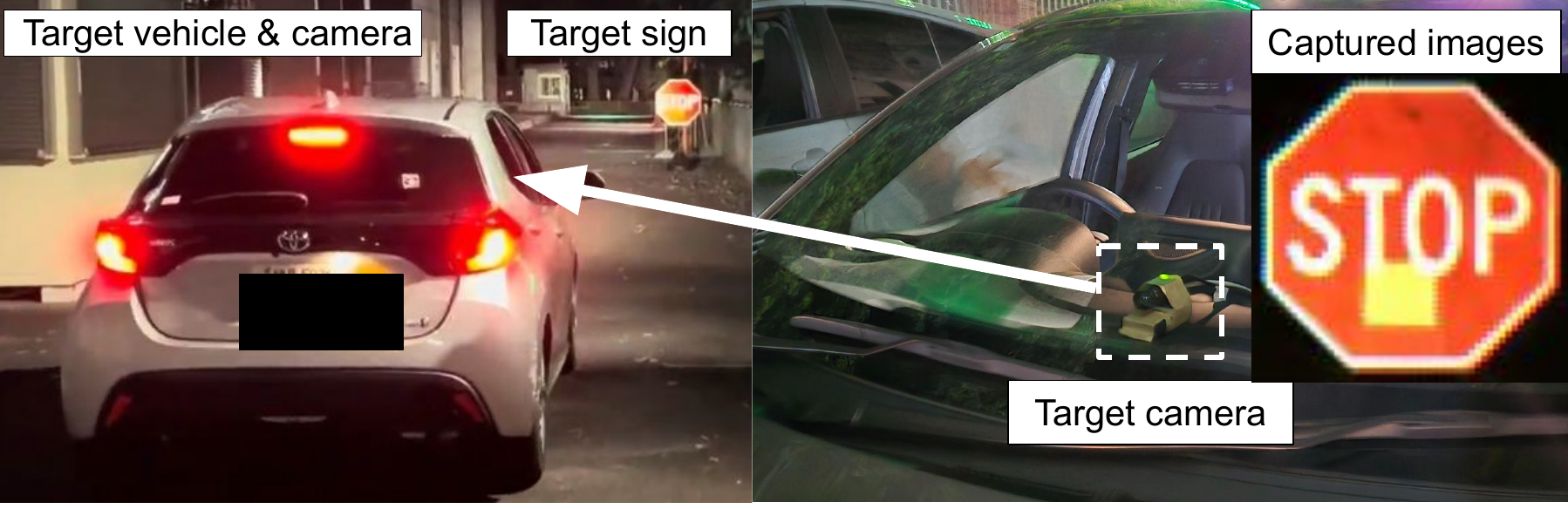}

    \caption{Experimental setup in a driving scenario. The camera is installed on the dashboard. The car drives at 5 km/h starting 50 m before the sign.}
    \label{fig:driving}

\end{figure}

\nsection{Experimental Setup for Driving Evaluation}\label{sec:driving-setup}
Figure~\ref{fig:driving} illustrates the experimental setup for the dynamic driving evaluation detailed in \S6.

\nsection{DPR Shield Technical Details}
    \label{appendix:dpr_shield}

     DPR Shield leverages the principle of light polarization, a fundamental physical phenomenon that describes the directional characteristics of electromagnetic wave oscillations. Naturally occurring light waves oscillate in all directions perpendicular to their propagation direction. A polarizing filter serves as an optical device that constrains these multi-directional oscillations to a single plane. Specifically, it transmits only the wave components that are parallel to its transmission axis while blocking all other components.
    
    When light is reflected from a surface, it generally maintains its incident polarization state. DPR Shield utilizes this property in conjunction with polarizing filters to develop an effective reflection control system. Our approach employs two polarizing filters strategically: one at the light source (headlights) to establish a controlled polarization state, and another at the camera to selectively modulate the intensity of reflected light~\cite{toshiba_teli_reflection_polarization}. This dual-filter configuration provides several key advantages in defending against retroreflective adversarial patches.
    
    For practical implementation, several considerations must be addressed. While the polarizing filters reduce the total amount of light reaching the camera, this can be mitigated through the use of higher intensity light sources in the headlights. Modern autonomous vehicle cameras often already incorporate polarization capabilities for managing reflections and glare~\cite{teledyne_polarized_sensor}, making the integration of DPR Shield highly practical. The system can even improve performance in low-light scenarios~\cite{hampson2024autonomous}. The effectiveness of the defense can be further optimized through careful consideration of filter placement and orientation relative to the vehicle's optical system.

        \begin{figure}[htb]
        \centering
        \includegraphics[width=\linewidth]{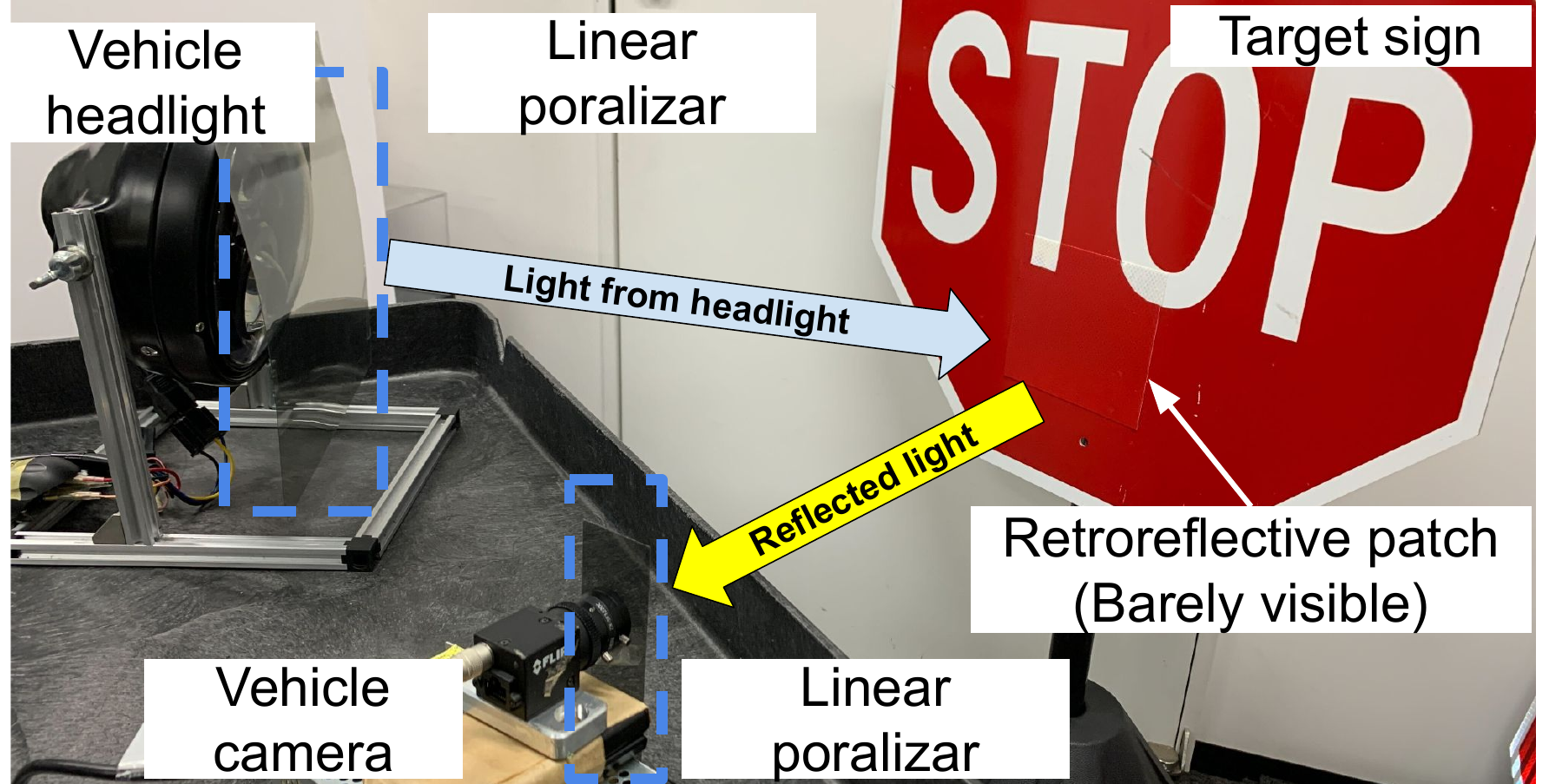}

        \caption{Setup of DPR Shield. Linear polarizer are attached to both the camera and headlight. These polarizer are also available as films that can be applied to surfaces of them.}
        \label{fig:defense-setup}
    \end{figure}

\nsection{Defense with Single Polarizing Filter}~\label{sec:single_filter}
    We evaluated the effectiveness of a single polarizing filter as a defense mechanism against ARP attack.

    \nsubsection{Experimental Setup}

    We conducted experiments using the same configuration described in \S\ref{sec:eval_real}, focusing on the scenario with the highest ASR: $d_{lon} = 15 \text{m}$, $d_{lat} = 0 m$. For attack configurations, we maintained the parameters that demonstrated maximum effectiveness in our previous evaluations. Specifically, for STOP signs, we used DG4090 material with $MPR$ of 0.1875 and 0.125 for single-stage and two-stage architectures, respectively. For SL65 signs, we employed NittoL material with $MPR$ of 0.0625 for both architectures.
    For a defense mechanism, we positioned a linear polarizing fileter vertically in front of the camera lens.
    We captured 100 images and measured the ASR with and without the filter to assess its defense capability.
    
    \nsubsection{Results}

    As shown in Table~\ref{tab:filter_effectiveness}, our results reveal that the effectiveness of a single-filter defense is limited and varies significantly by sign type.
    For STOP signs, the ASR decreased from 100 \% to 4\% for hiding attacks but only reduced to 89 \% for the altering attacks.
    In contrast, for SL65 signs, the ASR remains at 100 \% for both architectures, indicating single-filter defense has no defensive effect.
    There are two primary reasons for this inconsistency. First, a single filter reduces overall incoming light uniformly rather than selectively blocking only the reflected light from adversarial patches. Second, the optical characteristics of different colored patches play a crucial role—red patches on STOP signs require higher light intensity to maintain visibility and thus are partially suppressed by the filter's light reduction, whereas black patches on SL65 signs remain visually effective even with minimal light, allowing the attack to succeed despite the filtering.

\begin{table}[t]
    \centering
    \footnotesize
    \setlength{\tabcolsep}{1.5pt}
    \renewcommand{\arraystretch}{1.0}
    \caption{Comparison of ASR under one filter and without filters mounted in front of camera lenses. }
    \begin{tabular}{l cc cc}
        \toprule
        & \multicolumn{2}{c}{STOP} & \multicolumn{2}{c}{SL65} \\
        \cmidrule(lr){2-3} \cmidrule(lr){4-5}
        & Single & Two & Single & Two \\
        \midrule
        No Filter  & 100\% & 100\% & 100\% & 100\% \\
        With One Filter & 89\% & 4\% & 100\% & 100\% \\
        \bottomrule
    \end{tabular}
    \label{tab:filter_effectiveness}
    \vspace{-0.17in}
\end{table}

\nsection{Defense Effectiveness Evaluation of Existing Certifiable Patch Defenses}~\label{sec:patchcleanser-eval}

To evaluate the effectiveness of certified defenses against our ARP attack, we conducted experiments with PatchCleanser~\cite{xiang2022patchcleanser}, a state-of-the-art certified defenses against adversarial patch attacks in image classification.

\nsubsection{Experimental Setup}

We used the same models and attack configurations described in \S\ref{sec:eval_real}.
Specifically, we targeted both single-stage object detection models and two-stage classification models trained on the ARTS dataset. The attack configurations remained consistent with our previous experiments: for STOP signs, we used DG4090 material with MPR of 0.1875 and 0.125 for single-stage and two-stage architectures, respectively; for SL65 signs, we employed NittoL material with MPR of 0.0625 for both architectures.

We used PatchCleanser as a defense implementation, and followed the default configuration as specified in the original paper, with $\text{patch\_size} = 32$ and $\text{patch\_num} = 6$. We evaluated the classification accuracy under both benign (without attack) and ARP attack.

\nsubsection{Evaluation Metrics}
We use the No Defense Accuracy, Clean Accuracy, and Certified Accuracy following the PatchCleanser papers~\cite{xiang2022patchcleanser}.
No Defense Accuracy refers to the model accuracy without applying the PatchCleanser defense mechanism. Clean Accuracy measures the model accuracy when PatchCleanser is applied, indicating practical performance. Certified Accuracy represents the percentage of inputs for which PatchCleanser can provide a theoretical guarantee of correct classification regardless of patch attacks.
    \begin{table}
    \centering
    \footnotesize
    \setlength{\tabcolsep}{1.5pt}
    \renewcommand{\arraystretch}{1.0}
    \caption{Defense evaluation of PatchCleanser against ARP attacks. Clean Acc. is the rate of instances where PatchCleanser can correctly classified. Certified Acc means the rate of instances where PatchCleanser can certify.}
        \begin{tabular}{l cc cc}
        \hline
        \multirow{2}{*}{} & \multicolumn{2}{c}{Benign} & \multicolumn{2}{c}{Attack} \\
                \cmidrule(lr){2-3} \cmidrule(lr){4-5}
         & STOP & SL65 & STOP & SL65 \\
        \hline
        No Defense & 100\% & 100\%  & 0\%  & 0\%  \\
        
        Clean Acc & 0\%  & 100\%  & 0\%  & 0\%  \\
        
        Certified Acc & 0\%  & 0\%  & 0\%  & 0\%  \\
        \hline
        
        \end{tabular}
        \label{tab:PC_acc}
    \end{table}

    \nsubsection{Results}

    Table.~\ref{tab:PC_acc} show the results of our evaluation on both images with and without ARP ataccs.
    Our results reveal the limitations of PatchCleanser when applied to traffic sign recognition task under ARP attacks.
    For STOP signs, PatchCleanser not only cannot defend agains ARP attacks but even also reduce the accuracy of benign cases.
    Furthermore, PatchCleanser failed to provide any protection against our ARP attacks, with attack success remaining at 100\% (0\% accuracy) for both sign types.
    The certified accuracy of 0\% across all conditions indicates that PatchCleanser cannot provide any theoretical guarantees of robustness against our ARP attacks. These findings align with observations by Sato et al.~\cite{sato2024invisible} regarding the inherent limitations of masking-based certified defenses for traffic sign recognition. 
    The critical semantic information encoded in specific regions of traffic signs makes them particularly vulnerable to defensive strategies that rely on random masking. This evaluation underscores the need for alternative defense approaches that specifically address the unique challenges posed by ARP attacks on traffic sign recognition systems, motivating our development of the DPR Shield described in \S\ref{sec:defense}

    \nsection{ Distance Independence of the IoR Level}\label{sec:IoRlevel}
 
        We prove that the IoR level ($r$) remains effectively constant during vehicle movement. The IoR level is defined as:
        
        \begin{equation}
        r = \frac{\pi \cdot R'}{\cos\beta \cdot \cos\gamma}
        \end{equation}
        
        where $\beta$ is the entrance angle and $\gamma$ is the viewing angle.
        
        With the camera at origin $[0,0,0]$ and the sign at $[-d_{lon}, d_{lat}, h_s-h_l]$, the normal vector of the retroreflective patch is $[1,0,0]$ in the situation in Fig~\ref{fig:threat_model}. The vector from the patch to the camera is $[d_{lon}, -d_{lat}, -(h_s-h_l)]$, giving us:
        
        \begin{equation}
        \cos\beta = \frac{d_{lon}}{\sqrt{d_{lon}^2 + d_{lat}^2 + (h_s-h_l)^2}}
        \end{equation}
        
        \begin{equation}
        \cos\gamma = \frac{d_{lon}}{\sqrt{d_{lon}^2 + (d_{lat}-w_{car}/2)^2 + (h_s-h_l)^2}}
        \end{equation}
        
        We can show that $\cos\beta \approx \cos\gamma$ since the vehicle width is small compared to the distance. Through Taylor expansion and simplification, their relative difference is approximately:
        
        \begin{equation}
        \frac{\cos\gamma - \cos\beta}{\cos\beta} \approx \frac{d_{lat} \cdot w_{car}/2}{d_{lon}^2}
        \end{equation}
        
        With typical values ($d_{lat}=1.85\text{m}$, $w_{car}/2=0.9\text{m}$), the relative difference at 15m is less than 0.1\%, making $\cos\beta \approx \cos\gamma$ a valid approximation.
        
        Therefore, $r \approx \frac{\pi \cdot R'}{(\cos\beta)^2}$. Under practical conditions where $d_{lon} \gg d_{lat}$ and $d_{lon} \gg (h_s-h_l)$:
        
        \begin{equation}
        \cos\beta \approx 1 - \frac{\varepsilon^2}{2d_{lon}^2}
        \end{equation}
        
        where $\varepsilon^2 = d_{lat}^2 + (h_s-h_l)^2$, leading to:
        
        \begin{equation}
        r \approx \pi \cdot R' \cdot \left(1 + \frac{\varepsilon^2}{d_{lon}^2}\right)
        \end{equation}
        
        Our numerical analysis shows that in the practical detection range (15-50m), the IoR level varies by less than 2\%. At distances over 30m, the variation becomes negligible ($<$0.5\%). This confirms that the IoR level can be considered effectively constant during vehicle movement, especially at typical detection distances.

        \nsection{Detail of Retroreflective Material Characteristics and Color Analysis}\label{sec:detail-retro-analysis}

        We show the detail of the analysis of the retroreflective material characteristics and color analysis in Table~\ref{table:selected_patch_eval}.
        
        \begin{table*}[tbh]
\setlength{\tabcolsep}{15pt}
\renewcommand{\arraystretch}{1.1}
\centering
\footnotesize
\caption{Comparative analysis of retroreflective materials used in this study. The table shows color characteristics under daytime (600 lux) and nighttime (with 3400 lm headlight) conditions, along with color differences measured by L2 norm. }
\vspace{-0.1in}
\begin{tabular}{lcccc}
\toprule
Name in this paper & NittoL & HIP3930 & Nikkalite & DG4090 \\ \hline
Daytime (600 lux
) Color (RGB) & (110, 47, 25) & (86, 33, 17) & (93, 39, 24) & (94, 36, 18) \\
Nighttime with headlight (3400 lm) Color (RGB) & (255, 246, 80) & (252, 244, 156) & (255, 255, 189) & (255, 255, 254) \\
Day-Night Diff (L2 norm) & 251.83 & 302.53 & 316.44 & 359.78 \\
Diff from White (L2 norm) & 175.35 & 99.90 & 66.10 & 0.61 \\
Diff from STOP sign red (L2 norm) & 11.79 & 37.99 & 28.63 & 29.22 \\ \toprule
\end{tabular}
\label{table:selected_patch_eval}
\vspace{-0.15in}
\end{table*}